# A Nontrivial Interplay between Triadic Closure, Preferential, and Anti-Preferential Attachment: New Insights from Online Data

[1,2]Ivan V. Kozitsin, [3]Eduard R. Sayfulin, [3]Vyacheslav L. Goiko

[1]Laboratory of Active Systems, V. A. Trapeznikov Institute of Control Sciences of Russian Academy of Sciences, 65 Profsoyuznaya street, Moscow, 117997, Russian Federation

[2]Department of Higher Mathematics, Moscow Institute of Physics and Technology, 9 Institutskiy per., Dolgoprudny, Moscow Region, 141701, Russian Federation

[3]The Center of Applied Big Data Analysis, National Research Tomsk State University, 36 Lenina Street, Tomsk, 634050, Russian Federation

Email: kozitsin.ivan@mail.ru

**Abstract**

This paper presents an analysis of a temporal network that describes the social connections of a large-scale (~30,000) sample of online social network users, inhabitants of a fixed city. We tested how the main network formation determinants – transitivity, preferential attachment, and social selection – contribute to network evolution. Among other things, we found that opinion social selection does affect tie appearing whereas its impact on tie removing is rather unclear. We report that transitivity displayed the strongest effect on network dynamics. Surprisingly, a closer look revealed an intriguing and complex interplay between the transitivity, preferential attachment, and anti-preferential attachment mechanisms. For a given pair of unconnected nodes, if they have no mutual connections, then the probability of tie creation goes up with the total number of nodes' friends – that is exactly what the preferential attachment mechanism is assumed to do. Instead, if the nodes have at least one common friend, then the highest probability of tie appearing is achieved if both the nodes have only a few friends – a phenomenon that is called anti-preferential attachment. We attempted to explain this finding by appealing to the notions of social communities and leaders.

## 1. Introduction

Social networks establish paths along which information and social influence flow in our world. In the era of digitalization, it is extremely important to understand the key mechanisms behind network formation processes because this knowledge may help us to mitigate the harmful effects of echo chambers and misinformation spreading (Garimella et al., 2018; Jasny et al., 2015; Sasahara et al., 2021). To date, a huge scope of network formation models has been introduced, based on different empirically grounded mechanisms such as transitivity (aka triadic closure), preferential attachment, social selection, and others (Chandrasekhar & Jackson, 2016; Currarini et al., 2009; Das & Ghosh, 2021; M. E. J. Newman, 2001; Sendiña-Nadal et al., 2016; Yuan et al., 2018; Zhang et al., 2018). However, all these models still find it difficult to capture all the principal properties of real-world social networks, such as clustering, fat-tailed (power-law) degree distributions, homophily, small-worldness, and degree assortativity.

Because of this issue, there is a need for further empirical and theoretical research on social networks. This paper contributes to the first direction and presents a comprehensive analysis of a temporal network that describes the social connections of a sample of online social network users, citizens of a fixed city. The longitudinal data includes information on both opinion and non-opinion characteristics (age, gender), giving us an opportunity to study not only structural patterns of network evolution, but also pay attention to how nodal characteristics affect the network dynamics.

*1.1. Transitivity and Preferential Attachment*

Transitivity and preferential attachment are two, perhaps, the most established and widely reported determinants of network formation processes (Inoue et al., 2018; M. E. J. Newman, 2001; Peng, 2015; Zhang et al., 2018). Transitivity refers to the tendency of people to create connections with those with whom they have many friends in common. That is, the probability of tie appearing between two nodes should go up with the number of friends the nodes share. The transitivity mechanism is a natural explanation of the persistent clustering of social networks (M. E. J. Newman, 2001). In turn, the preferential attachment rule postulates that we are more likely to form ties with popular individuals – those who have many social connections. According to the preferential attachment rule, the probability of tie creation should go up with the nodal degree. It is worth noting that the preferential attachment mechanism, conceptualized by the prominent Barabasi-Albert model, generates synthetic networks with scale-free degree distributions (Albert & Barabási, 2002). Importance of the Barabasi-Albert model stems from the fact that real social networks tend to have the same property – social networks are typically scale-free.

*1.2. About homophily and selectivity*

This paper explores both the static and dynamic properties of the networked system. In this regard, we will distinguish between the concepts of *homophily* (a static feature) and *selectivity* (dynamic feature), which are often used interchangeably in research. Let us now explain our notation approach in more detail.

If we say that a network $G$ is homophilic (or assortative) with respect to some attribute $x$, it means that we take a look on $G$ at a fixed time moment $t$ – that is, observe its snapshot $G(t)$, – and see that $G(t)$'s neighboring nodes tend to be alike in terms of this attribute $x$. A possible approach to understanding whether $G(t)$ is homophilic with respect to $x$ is to calculate the assortativity coefficient $C(G(t), x) \in [-1,1]$ (its detailed definition can be found in Appendix A). In a nutshell, the assortativity coefficient can be understood as a sort of the Pearson correlation coefficient that checks whether neighboring nodes correlate in terms of the attribute by comparing the current network's configuration against a network built upon the same nodes connected by the same number of edges but placed purely at random (the null model). For homophilic networks, the assortativity coefficient takes the positive values. Empirical studies suggest that social networks tend to be homophilic with respect to a broad spectrum of nodal characteristics (age, gender, culture, religion, opinions, etc.) (Halberstam & Knight, 2016; McPherson et al., 2001; M. E. J. Newman, 2003; Noldus & Van Mieghem, 2015).

In turn, we will refer to selectivity as the tendency of individuals to (i) create new ties with those people who have similar attributes and (ii) delete social connections if these connections bring together individuals with too dissimilar characteristics. Typically, scholars pay more attention to the first facet of selectivity, ignoring the process of tie removing. Nonetheless, analysis of real-world temporal social networks revealed that selectivity is a persistent feature of network formation processes. Similarly to homophily, one can emphasize age, gender, culture, religion, and opinion social selection (Lewis et al., 2012; Steglich et al., 2010; Zhang et al., 2018).

Because of this issue, one may assume that network homophily stems from social selection. This is partly true because not only selectivity may contribute to network assortativity (Holme & Newman, 2006). One prominent example is social influence, which causes people to change their minds and reduces disagreement along social ties. Because of social influence, opinions of neighboring nodes become closer to each other, and thus the resulting network features a higher level of opinion homophily. There is no consensus in the scientific community regarding how strong social influence and opinion selectivity contribute to opinion homophily. Some studies witness the prevalence of social influence (Lazer et al., 2010), others speak in favor of selectivity (Lewis et al., 2012), whereas Ref. (Wang et al., 2020) observed no signs of both peer influence and social selection. In turn, Ref. (Chang, 2022) reported that both selection and influence can prevail depending on what type of academic behavior is under consideration. Apparently, context is important here, so depending on empirical settings, the effects of selectivity and social influence may have different strengths.

### *1.3. Contributions*

Our contributions are as follows. First, we built a new dataset that represents the coevolution of political opinions and social ties of a large sample (~30,000) of online social network users living in the same city, according to the information from their profiles. Despite the fact that we used this data to study network formation processes only, the dataset may be of particular interest for scholars who work in the field of opinion dynamics models.

We revealed a clear asymmetry in the structure of opinion homophily: whereas only individuals with radical opinions tend to have more connections to those of similar positions than one should expect by chance, one side of opinion spectrum features a remarkably higher tendency to maintain ideologically homogeneous networks than its ideological counterpart. Importantly, for elderly individuals, only the representatives of this side of opinion spectrum do demonstrate assortative mixing whereas other opinion groups do not.

Next, we observed a clear effect of social selection with respect to both opinion and non-opinion (age, gender) attributes on how new ties appeared in the network. Although the influence of opinion selectivity on the probability of tie creation was not as strong as those of age or gender selectivity, it was valid even after controlling for other covariates. Interestingly, we demonstrated that age-related social selection, which played a crucial role in tie creation, does not affect edge removing. The effect of opinion selectivity on edge removal is rather unclear.

As opposed to social selection mechanisms, transitivity displays the most prominent influence on network evolution. Surprisingly, a closer look revealed an intriguing and complex interplay between the transitivity, preferential attachment, and anti-preferential attachment mechanisms. We proved that the first two of them, which are, perhaps, the most famous network determinants, are mutually dependent and do not operate simultaneously: for a given pair of unconnected nodes, if they have no mutual connections, then the probability of tie creation goes up with the total number of nodes' friends. Instead, if the nodes have at least one common friend, then the highest probability of tie appearing is achieved if both the nodes have only a few friends – anti-preferential attachment is in charge. Interestingly, this peculiarity does not work in reverse: the probability of tie removing goes down with the total number of connected nodes regardless of the number of common peers, so the anti-preferential attachment does not affect tie removing. We attempted to explain these findings by appealing to the notions of social communities and leaders.

**2. Brief review of the VKontakte organization**

VKontakte (VK) is a Russian online network whose organization can be briefly presented as follows. First, it includes two main types of accounts: ordinary (native) users and information sources. The latter are typically users with a large number of followers (more than or equal to 1000 – so-called bloggers) or public pages (in a typical case, a public page is designated to disseminate some information, for example, it could be a representative of a mass media outlet). Ties could appear between VK accounts, both one-directional (a user follows another user, another user follows an information source) and two-directional (between two users), which is commonly referred to as friendship connections. A friendship tie is formed when a user $i$ starts to follow a different user $j$ (on this occasion, $j$ receives an $i$'s friend request), and $j$ replies to this request by starting to follow $i$. While new ties appear between VK users, some ties can also be removed. To remove a friendship connection, one of two friends should click the button "cancel a subscription." The same button cancels a one-directional tie.

To summarize, each VK user is characterized by two sets of accounts: friends and followees. The former are usually native users (however, sometimes bloggers can also establish friendship-type connections, but such situations are rare). The latter typically include bloggers, public pages, and (less frequently) native users (who have not responded to friend requests for some reason). Of course, there are other types of accounts: groups and events, but these accounts contribute less to information dissemination, so we ignore them in our study.

The architecture of VK can be safely approximated by a bipartite graph $G = (U, IS, E_U, E_{U \to IS})$, with $U$ describing the nodes of native users and $IS$ standing for information sources' vertices. The set $E_U$ includes ties between the nodes from $U$ (these ties are usually bidirectional, less frequently – one-directional), while ties from the set $E_{U \to IS}$ are directed from the $U$ vertices to those from the set $IS$.

Information that users receive on VK is formed by the news feeds that display their friends' and followees' published posts, comments, likes, etc. Of course, this data is curated by ranking algorithms, which can drastically alter the initial (time-based) order by giving more weight to content that has a higher chance of being appreciated by users.

We did not mention online bots in the description of the VK population (Pescetelli et al., 2022). These specific accounts hold rather an intermediate position in our classification: on the one hand, they are not native users and, importantly, the friendship (and following-type) connections between ordinary users and bots are rare. On the other hand, bots cannot be classified as meaningful information sources. A more precise categorization would be to define online bots as satellites of information sources because their best-known behavior strategy is to comment information sources' posts and, by doing so, influence native users.

## 3. Data and Notations

Our analysis started with a sample of $N = 29,248$ VK users. These individuals were chosen among all the Russian-located VK users as those whose accounts were not closed by privacy settings, active (i.e., visited no less than once per month), and who indicated the same Russian city (its name is not disclosed to ensure the anonymity of the data) as the place of residence. This geographical restriction partially excluded the confounding factor of geographical proximity between users, which is an important determinant of social network evolution processes (Lewis et al., 2012). The city was chosen as the capital of a Russian region, that is sufficiently distant from other large cities – this criterion facilitated the closeness of the social system. Furthermore, to ensure the precision of our estimations, a city had to be as large as our computing power could handle. It is worth noting that there was a list of cities that meet our criteria, some of them were used to test the generalizability of our results (see Section 7). This geographical restriction facilitates the closeness of the social system. Additionally, each user was older than 17 and the number of their followees was in the interval of $[10, 200]$. The latter criterion makes our opinion estimation algorithm (see below) more precise. Besides, the sample was cleaned of bloggers because we used them in opinion evaluation. Without this step, opinion estimations would not be independent from the social network. In this regard, the sample was supposed to include only native VK users.

Next, for these users, two social graphs were constructed based on their friendship connections: at times $t_1$ (around March 2019) and $t_2$ (around September 2019) – we refer to them as to networks $G(t_1)$ and $G(t_2)$ correspondingly (the underlying adjacency matrices are denoted by $A(t_1) = \left[a_{ij}(t_1)\right]_{i,j=1}^{N}$ and $A(t_2) = \left[a_{ij}(t_2)\right]_{i,j=1}^{N}$, where $a_{ij}(t)$ signifies whether there is a link between vertices $i$ and $j$ at time $t$). Pilots revealed that a time span of six months is sufficiently large to detect notable changes in the social graph. Further, at this time scale, the majority of social graph changes have a relatively atomic character (less than one new link and one removed link per user).

Due to some advanced privacy settings, a user can hide the list (or some) of their friends even though the general information on their account is available. However, these missing data can be restored "from the other end" (but only if a "hidden" friend does not hide their friend list as well) – by symmetrizing the adjacency matrices $A(t_1)$ and $A(t_2)$. In the resulting (symmetric) social graphs $G(t_1)$ and $G(t_2)$, the giant connected components were identified that included 27,861 (95% of all nodes) and 28,055 (96% of all nodes) nodes, correspondingly. Note that each node from the giant connected component of the graph $G(t_1)$ belongs also to the giant connected component of the graph $G(t_2)$. On this basis, we decided to focus on the giant connected component of the graph $G(t_1)$. After fixing its vertices, we obtained two new graphs (for simplicity, we still call them $G(t_1)$ and $G(t_2)$) built upon the same sample of users. Note also that while $G(t_1)$ is a connected graph, the graph $G(t_2)$ is disconnected and, in addition to a giant component, includes 42 nodes with no friends (single-node components).

Our sampling strategy as well as our approach to focusing on the giant connected component, among other things, reduced the probability of encountering online bots because they tend to occupy the periphery of online networks and, thus, they were likely to be removed from the sample after discarding isolated small connected components (González-Bailón & De Domenico, 2021).

For each user $i$ from the sample, their age ($age_i$), gender ($gender_i$), and the set of $i$'s followees were downloaded from VK. The information on users' followees was downloaded twice: at times $t_1$ and $t_2$. The set of $i$'s followees at time $t$ is denoted by $S_i(t)$ (its cardinal number is signified by $s_i(t) = |S_i(t)|$). The user $i$'s friends at time $t$ is symbolized by $F_i(t)$ (with the cardinal number $f_i(t) = |F_i(t)|$). It is worthy to note that due to our sampling methodology, for each $i$, variables $age_i$ and $gender_i$ are defined, the sets $f_i(t)$ and $s_i(t)$ are nonempty (at both the time moments), and the following inequalities hold: $age_i \geq 18, f_i(t) > 0$, and $10 \leq s_i(t) \leq 200$.

Further, for each user $i$, we calculated the following sets: (1) $i$'s new friends $F_i^{new} = \{j \mid j \in F_i(t_2), j \notin F_i(t_1)\}$, (2) $i$'s deleted (removed) friends $F_i^{del} = \{j \mid j \notin F_i(t_2), j \in F_i(t_1)\}$, (3) $i$'s new followees $S_i^{new} = \{j \mid j \in S_i(t_2), j \notin S_i(t_1)\}$, (4) $i$'s deleted followees $S_i^{del} = \{j \mid j \notin S_i(t_2), j \in S_i(t_1)\}$. The cardinal numbers of these sets are denoted by $f_i^{new}, f_i^{del}, s_i^{new}$, and $s_i^{del}$ correspondingly.

Based on the information on the users' subscriptions to information sources, we estimated their opinions on a political issue using the methodology from Ref. (Kozitsin et al., 2020). In a nutshell, the opinion estimation algorithm, which is based on the logit model, projects the set of the user $i$'s information sources $S_i$ into the interval $[0,1]$, with 0 corresponding to the conservative political stance (supporting President Putin or leaders of the Russian systemic opposition) and 1 standing for the support of the Russian non-systemic opposition (see Appendix B for details). Because we had two snapshots of users' followees, we obtained two opinion estimations for each user. In the rest of the paper, the user $i$'s opinion at time $t$ is signified by $x_i(t)$. For simplicity of notation, users that espouse opinions located at the left side of opinion spectrum are called conservatives, while those whose opinions are near the right edge of the interval $[0,1]$ are referred to as liberals. Individuals with $x(t) \approx 0.5$ are labelled as moderates (we recognize that such designations are incorrect according to political science theory).

The data are two snapshots ($G(t_1)$ and $G(t_2)$) of a temporal network that describes friendship connections between 27,861 VK users—citizens of a city. Each snapshot also contains information on user age, gender, and followees (subscriptions to VK information sources – public pages and bloggers) attributes. Based on users' followee lists, we estimated users' opinions on a political scale (estimated opinions lie in the interval $[0,1]$) and thus advanced the data with two opinion snapshots. All the data were gathered and proceeded in an anonymous fashion, without storing and revealing any private information.

## 4. Preliminary Analysis

### *4.1. The baseline graph properties*

In Table 1, we present the baseline properties of two network snapshots. In line with the empirical studies on social networks (M. Newman, 2018), the graph snapshots exhibit power-law degree distribution (see Figure 1), positive clustering (captured by two measures – the transitivity coefficient and the average clustering coefficient), and assortative mixing (with respect to nodal degree, opinion, age, and gender attributes). Interestingly, different nodal attributes demonstrate various assortativity levels – age homophily takes a dominant position (with the value of ~0.49). The second largest homophily rate is characterized by gender homophily (~0.36) while the lowest value of the assortativity coefficient is observed for degree homophily (only ~0.12). As it can be seen from Table 1, all changes in the macroscopic metrics between times $t_1$ and $t_2$ are tiny.

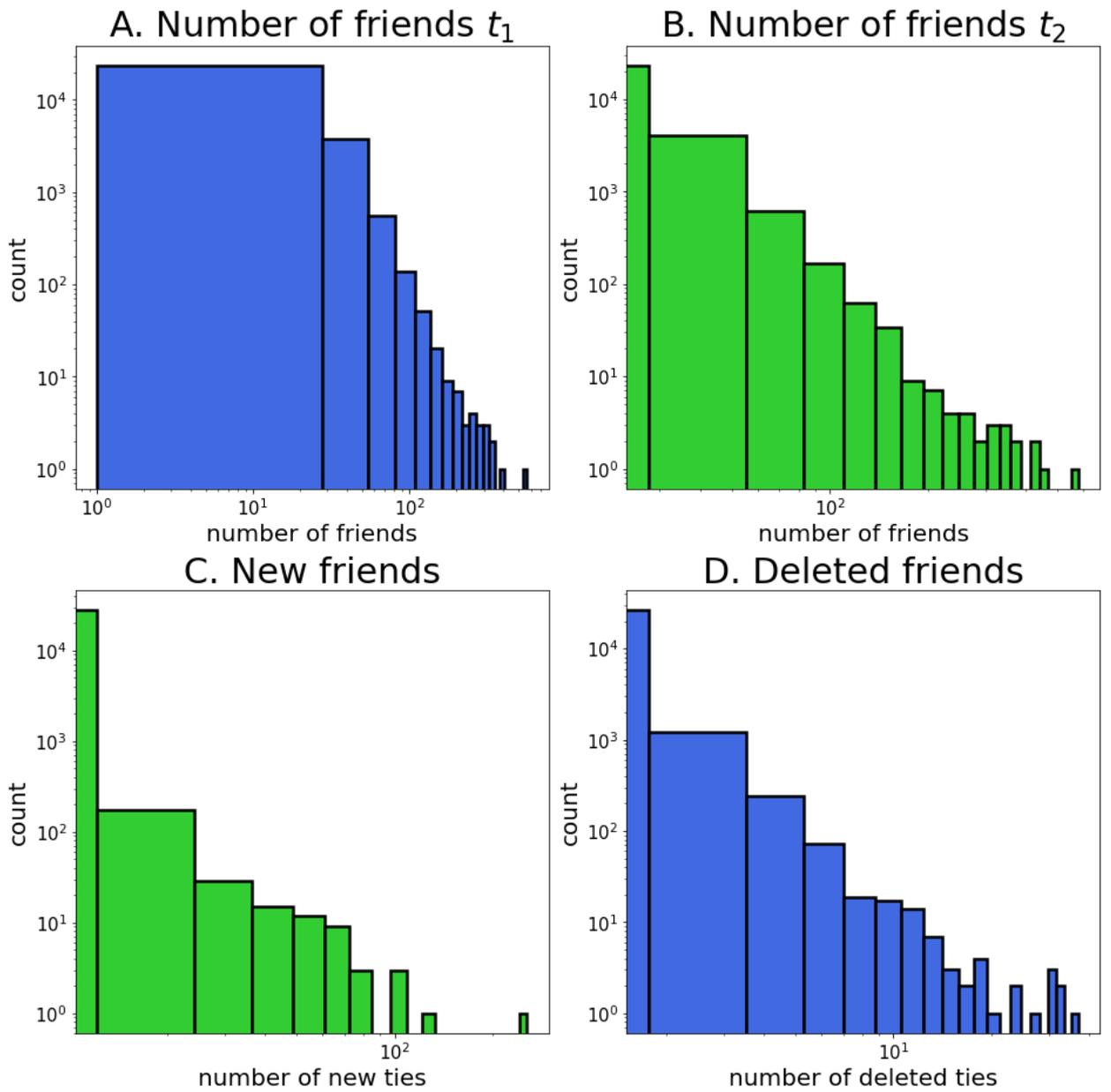

*Figure 1.* Vertex degree distributions at times $t_1$ and $t_2$ (upper panels). Distributions of newly appeared and deleted friends (bottom panels).

Table 1

Characteristics of graphs $G(t_1)$ and $G(t_2)$.

|  | $t_1$ | $t_2$ |
| --- | --- | --- |
| Density | 0.001 | 0.001 |
| Transitivity | 0.07 | 0.071 |
| Average clustering | 0.11 | 0.11 |
| Degree-based assortativity | 0.115 | 0.117 |
| Opinion-based assortativity | 0.226 | 0.233 |
| Age-based assortativity | 0.493 | 0.485 |
| Gender-based assortativity | 0.359 | 0.362 |

In the social system under consideration, the mean number of friends at time $t_1$ was 16.44, while the maximal number was 566 – there is only one such node. The second largest hub had only 384 friends, the third – 348 ones. Between times $t_1$ and $t_2$, 20,245 new ties appeared (1.45 new friends per user), and 4,562 ties were removed (0.33 deleted friends per user)[1] – that is, the graph slowly becomes denser. Along with these changes, the three hubs mentioned before increased the number of their friends from 566, 384, and 348 to 580, 417, and 380 correspondingly – the popular users became more popular. In turn, Figure 2 demonstrates that the sample is biased towards younger and more conservative individuals, while it is still gender-balanced.

---

[1] Note that our time resolution does not give the opportunity to identify the initiators of ties appearing and removing.

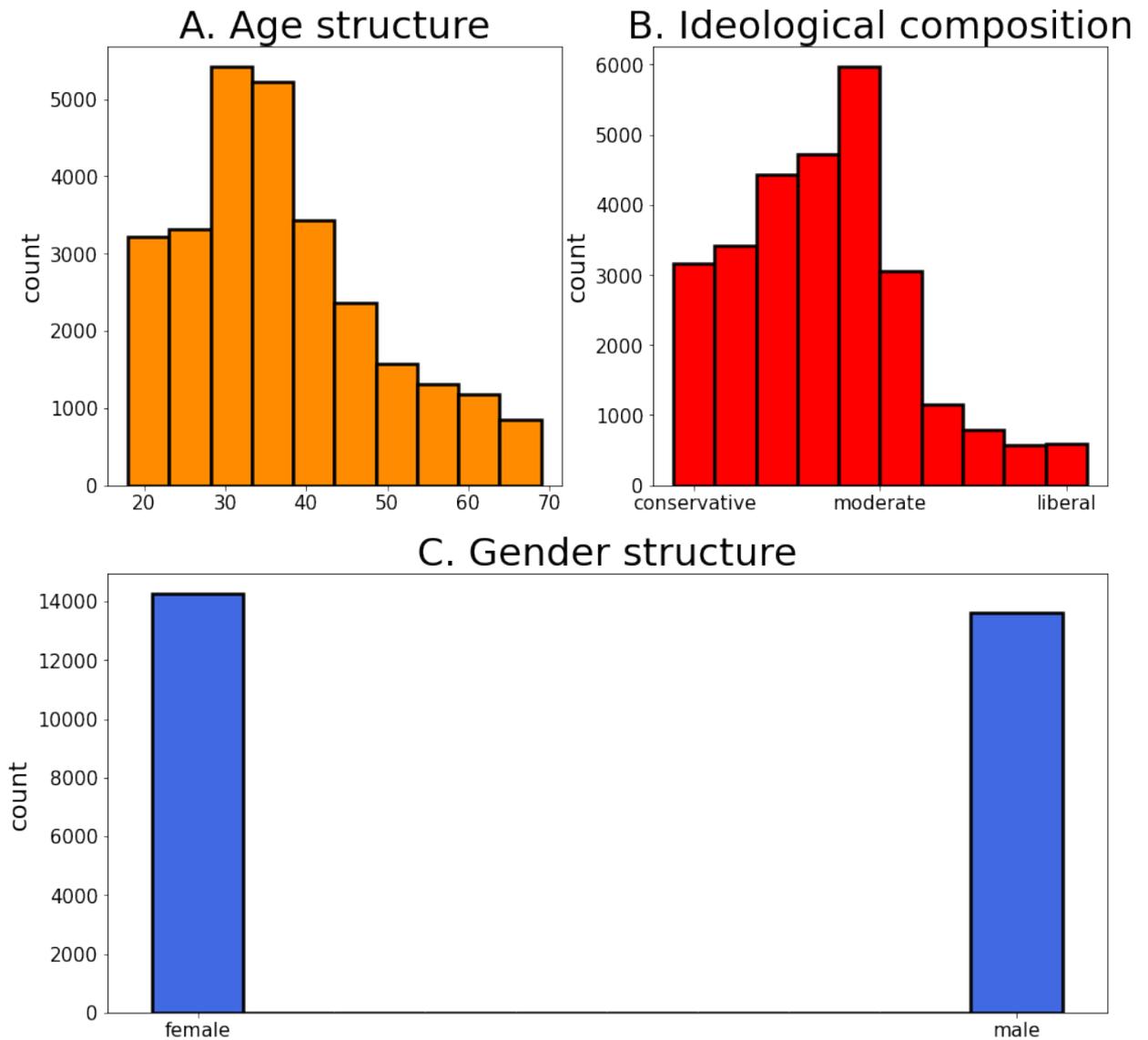

*Figure 2.* Age, opinion, and gender structure at time $t_1$.

*4.2. Correlation analysis*

We performed correlation analysis to discern causal relations between nodal characteristics (see Table 2). We found that younger individuals tend to be more liberal, have on average more friends and followees, and adjust their ego networks more often. We also obtained that users with more friends organize more new friendships and following-type connections, as well as tend to have more followees. Interestingly, the number of friends has extremely weak correlations with the number of new and deleted followees; this appears strange because those with more friends appear to be more active on VK and thus should renew their subscription lists more actively (even subject to our total number of followees restriction). However, these two quantities ($s^{new}$ and $s^{del}$) correlate positively with the total number of followees. We also report that males tend to have more friends on average but have fewer followees on average. Females also renew their followers more frequently. More detailed analysis revealed that the numbers of new and deleted friends depend positively and monotonically on the total number of friends, with no special activity patterns for low-degree vertices.

Table 2

Correlation matrix

|  | $age$ | $gender$ (1 – females, 2 – males) | $x$ | $f$ | $f^{new}$ | $f^{del}$ | $s$ | $s^{new}$ | $s^{del}$ |
|---|---|---|---|---|---|---|---|---|---|
| $age$ |  |  |  |  |  |  |  |  |  |
| $gender$ | −.17 *** |  |  |  |  |  |  |  |  |
| $x$ | −.2 *** | −.01 |  |  |  |  |  |  |  |
| $f$ | −.14 *** | .09 *** | .05 *** |  |  |  |  |  |  |
| $f^{new}$ | −.05 *** | .01 *** | .02 ** | .44 *** |  |  |  |  |  |
| $f^{del}$ | −.17 *** | −.02 ** | .06 *** | .27 *** | .18 *** |  |  |  |  |
| $s$ | −.13 *** | −.13 *** | −.3 *** | .12 *** | .06 *** | .09 *** |  |  |  |
| $s^{new}$ | −.05 *** | −.09 *** | −.05 *** | −.01 * | .18 *** | .07 *** | .26 *** |  |  |
| $s^{del}$ | −.12 *** | −.05 *** | .01 * | .03 *** | .07 *** | .13 *** | .33 *** | .19 *** |  |

*Note1*. To estimate pairwise correlations between $gender$ and other variables, we used the point biserial correlation coefficient

*Note 2*. Significance codes: 0 '***' 0.001 '**' 0.01 '*' 0.05 ' ' 1

### 4.3. Homophily organization

Previous studies revealed that in assortative networks, not all nodes contribute to assortativity equally, and it could be the case that mixing patterns may vary with the values of a specific nodal attribute (Noldus & Van Mieghem, 2015). To gain a better understanding of homophily organization, we divided the users into groups for each nodal characteristic, so that users within the same group had similar values of the focal characteristic, and then we computed the average compositions of users' neighborhoods across these groups (with respect to the focal characteristic). Such divisions give an opportunity to explore how the local patterns of homophily depend on the properties of the nodes. In particular, the continuous opinion spectrum [0,1] was discretized into five subintervals (ideological groups): 0–0.2 (Strong Conservatives – SC), 0.2–0.4 (Conservatives – C), 0.4–0.6 (Moderates – M), 0.6–0.8 (Liberals – L), and 0.8–1 (Strong Liberals – SL). We discovered that (i) liberal individuals have more friends and delete friends more frequently; (ii) individuals with radical political positions have more followers and (especially strong liberals) renew them more frequently after analyzing how users' properties differ across these ideological groups (see Table C1 in Appendix C). Further, we report that (iii) the population of radical individuals (Strong Liberals and Strong Conservatives) is slightly biased towards males. All these results were supported by the Mann-Whitney U test.

Let us now describe the structure of homophily in the system. We start with political (opinion) homophily. In line with previous empirical studies on political homophily (Barberá, 2014; Kozitsin, 2021), Figure 3 clearly demonstrates that more radical individuals are more prone to maintain ideologically-consistent connections while moderate individuals allocate their friendship ties almost as much as the null model predicts. Among radical users, Strong Liberals demonstrate a sounder tendency towards assortative mixing. Interestingly, these homophily patterns vary with user age: for older individuals, only Strong Liberals feature (extremely sound) assortativity mixing, while the other ideological groups' neighborhoods coincide with what the null model predicts. The tendency of Strong Liberals to maintain more ideologically consistent networks is observed for both females and males (see Figure C1 in Appendix C). To complement our analysis of opinion homophily, we took a look at those users that maintain ideologically heterogeneous neighborhoods. We found only five individuals in the sample who have connections to both Strong Conservatives and Strong Liberals at time $t_1$ (at time $t_2$, it was the same group of people). The minimal value of the estimated opinions among these users is $x = 0.45$, while their ages lie in the interval 32–35. Strictly speaking, we cannot refer to such users as to gatekeepers[2] (because we do not know what content they share with their friends). A more precise definition for such users would be "users with ideologically heterogenous ego-networks." Despite their population being extremely small, the Mann-Whitney U test indicates that such users are more liberal and tend to have more friends than other users from the same age group.

---

[2] Typically, gatekeepers are defined as users with access to content coming from sources of diverse opinions and who attempt to cover different points of view in their messages.

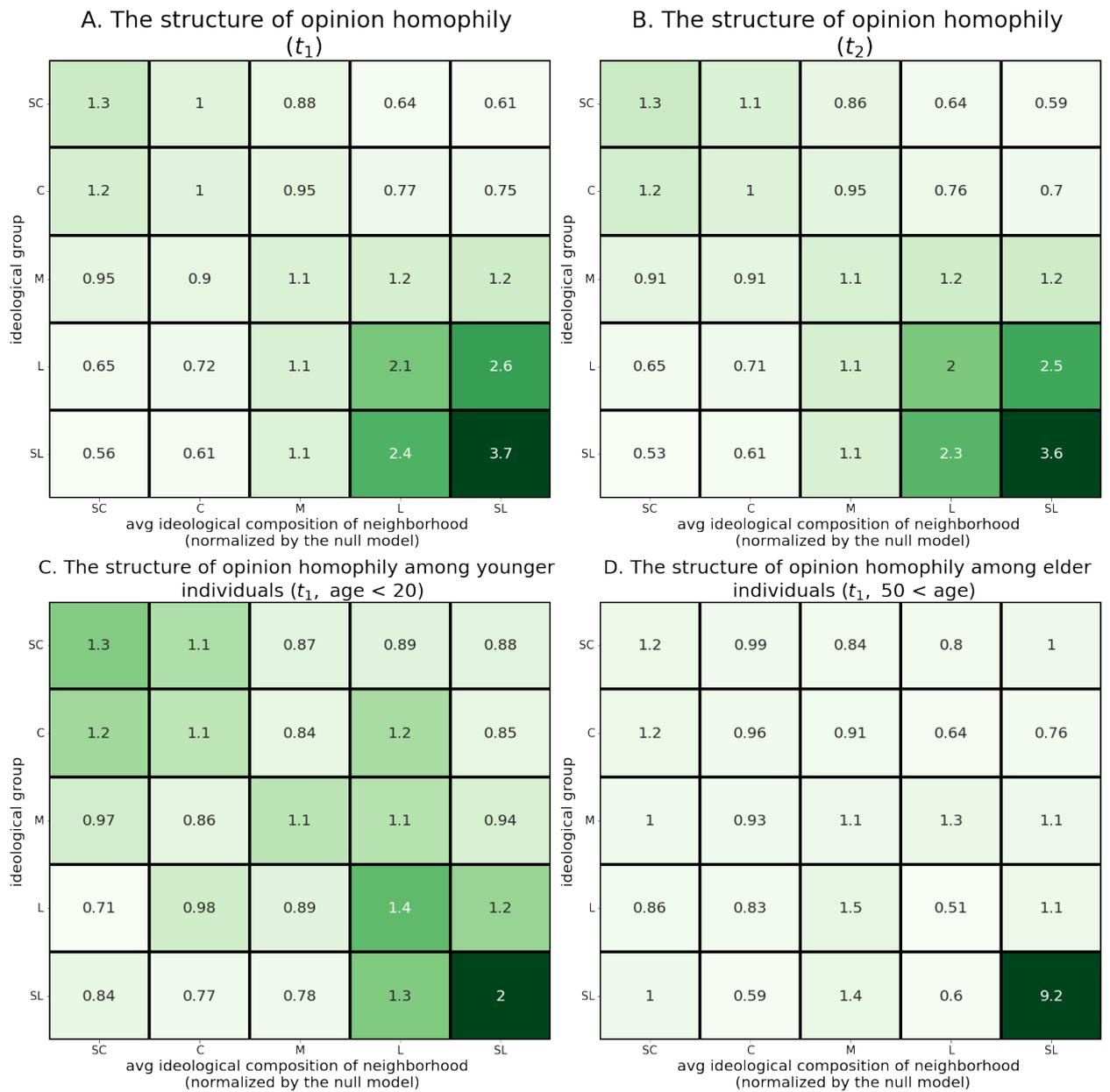

*Figure 3.* The structure of opinion homophily at times $t_1$, $t_2$ (panels A and B) and across two different age groups at time $t_1$ (panels C and D).

In general, age and gender homophily do not vary at the group level (see Figures C2 and C3 in Appendix)—all groups show a clear tendency to have more ties with individuals from the same group than the null model predicts. Some asymmetries can be found in the structure of age homophily, with younger users featuring a sounder tendency towards assortative mixing, compared to older age groups.

The organization of degree homophily is more interesting (see Figure 4). In social network theory, nodes with similar degrees should have more chances of being friends (M. E. J. Newman, 2003). Further, empirical studies reported that degree assortativity may vary with nodal degree (Noldus & Van Mieghem, 2015). In particular, positive values of degree assortativity may be achieved due to high-degree nodes, while other nodes may contribute to the overall degree assortativity in a negative fashion. We report that it is exactly our case. Our results clearly indicate that only high-degree users demonstrate a higher probability of being mutually connected than the null model predicts while the neighborhoods of users characterized by lower values of nodal degree are shifted substantially towards hubs.

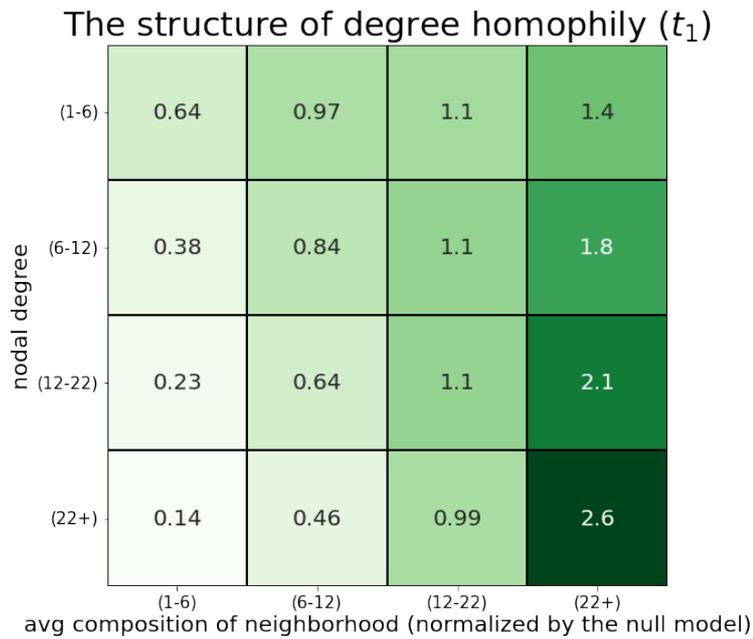

*Figure 4.* The structure of degree homophily at time $t_1$.

## 5. Analysis of the network dynamics

### 5.1. Overview of our approach

In this section, we identify the factors that govern network formation processes. Because these processes are governed by complex nonlinear effects whereby the independence assumptions are frequently violated (because social networks are highly correlated structures, as was demonstrated in previous empirical studies and in the previous section), the standard statistical approaches are not applicable here. A perfect tool to analyze longitudinal networked data is so-called stochastic actor-oriented models (SAOM) (Snijders, 2001, 2017). However, we failed to apply the SAOM-based software package RSiena[3] to our network, apparently due to the large size of the network. Because of this issue, we used the probabilistic approach (M. E. J. Newman, 2001) and calculated the probabilities of tie appearing and being removed separately, as functions of the properties of nodes. It is worthy of note that this approach requires a huge number of observations to ensure the precision of estimations. Notably, a similar approach is frequently implemented in link prediction tasks whereby various predictive models are used to forecast whether two disconnected vertices will become tied in the next observation time or whether two connected vertices will become disconnected in the next network snapshot (Kumar et al., 2020).

Our network includes $N = 27,861$ nodes. It implies $N * (N - 1)/2 = 388,103,730$ distinct pairs of vertices. At time $t_1$,

$$\sum_{i,j=1}^{N} a_{ij}(t_1) / 2 = 229,066$$

of these pairs were connected and, correspondingly,

$$\sum_{i,j=1}^{N} \left(1 - a_{ij}(t_1)\right) / 2 = 387,874,664$$

of them were disconnected. A straightforward way to estimate the probability of tie appearing is to compute the following proportion:

$$\frac{\sum_{i,j=1}^{N} \left(1 - a_{ij}(t_1)\right) * a_{ij}(t_2) / 2}{\sum_{i,j=1}^{N} \left(1 - a_{ij}(t_1)\right) / 2},$$

in which the nominator represents the number of newly formed edges between times $t_1$ and $t_2$. Correspondingly, the probability of tie removing can be presented as follows:

---

[3] https://www.stats.ox.ac.uk/~snijders/siena/

$$\frac{\sum_{i,j=1}^{N} a_{ij}(t_1) * \left(1 - a_{ij}(t_2)\right)/2}{\sum_{i,j=1}^{N} a_{ij}(t_1)/2}.$$

Next, we emphasize edge and nodal independent variables that characterize each pair of vertices in the network (even if they are not connected). The edge independent variables include, for example, the number of common friends, the number of common followees, or the absolute difference in opinions, age, etc. In turn, the nodal independent variables are presented by different pairwise combinations of nodal attributes, such as age, gender, or the nodal degree. For example, in the case of the gender attribute, each pair of vertices belongs to one of three types: "female-female", "female-male", or "male-male". For each of the opinion, age, and nodal degree attributes, we divide the range of its values into three subintervals to emphasize groups that are characterized by the similar values of the attribute. For the age attribute, the groups are (18-31), (31-40), and (40+). The nodal degree attribute is presented by the groups (1-8), (8-17), and (17+) correspondingly. In turn, the opinion attribute is characterized by the groups "cons", "mod", and "lib" that symbolize node opinions as follows. Cons: (0-0.33), mod: (0.33-0.66), lib: (0.66-1). Note that the age and nodal degree groups were separated by 33-rd and 66-th percentiles of the corresponding distributions.

We also would like to emphasize that we use new group stratifications here, different from those that were applied in the previous section. For example, previously, we employed five opinion groups ("SC", "C", "M", "L", "SL). However, such a division results in 15 pairwise combinations ("SC-SC", "SC-C", …, "SL-SL"). Pilots revealed that this division strategy will result in unneeded statistical fluctuations. So we decided to reduce the number of groups for each attribute up to three.

Then, we can compute the probabilities of tie appearing and disappearing for different values of (both edge- and nodal-based) independent variables (that is, we can compute estimated conditional probabilities). For example, the probability of tie appearing if the number of common friends is equal to zero can be obtained as:

$$\frac{\sum_{i,j=1}^{N} 1_{[CommFr_{ij}=0]} \left(1 - a_{ij}(t_1)\right) * a_{ij}(t_2) / 2}{\sum_{i,j=1}^{N} 1_{[CommFr_{ij}=0]} \left(1 - a_{ij}(t_1)\right) / 2},$$

where $1_{[\ldots]}$ denotes the event indicator function. One more illustrative example: the probability of tie appearing if one of two nodes espouse liberal views and the second one – conservative stance is given by the following expression:

$$\frac{\sum_{i,j=1}^{N} 1_{[LibCons]} \left(1 - a_{ij}(t_1)\right) * a_{ij}(t_2) / 2}{\sum_{i,j=1}^{N} 1_{[LibCons]} \left(1 - a_{ij}(t_1)\right) / 2},$$

where the event $LibCons$ is defined as:

$$LibCons = \{(x_i < 0.33) \& (x_j > 0.66)\} | \{(x_j < 0.33) \& (x_i > 0.66)\}.$$

Because we pay special attention to how opinion selectivity affects network formation, the probabilities of tie appearing and removing are analyzed not only in terms of nodal opinions (a nodal independent variable), but also using the absolute difference in nodes' opinions (an edge independent variable). While doing so, we already use four opinion groups: (0-0.25), (0.25-0.5), (0.5-0.75), (0.75-1) that are signified as "small", "avg", "large", and "huge" opinion differences. Whereas the nodal approach is more accurate and gives the opportunity to characterize how the dependent variable varies across the opinion groups (discerning some nontrivial group-specific dependencies), the edge approach takes a more general look at whether the opinion selectivity mechanism affects network evolution. Besides, the edge approach leads to more precise estimations with the same amount of data just because it implies fewer group divisions (four groups versus six ones in the case of the node-based approach).

For ease of presentation, the probabilities of tie appearing that we computed (which are actually extremely small) are displayed after multiplication by the factor of 1,000,000.

It is also important to note that the network evolution may be seriously affected by the opinion dynamics of nodes. Because the opinions of nodes are frequently changed – between times $t_1$ and $t_2$, 3,768 nodes (~14% of all vertices) changed their opinions by more than 0.05 (the value of 0.05 is used to get rid of noise opinion shifts) – one may expect that some edges appeared or were removed after the adjacent nodes changed their opinions. Such observations could seriously disrupt our analysis – it could be the case, for example, that opinions become closer to each other first, and then a tie between the underlying nodes appeared because the nodes' opinions are close enough by now. Because of this issue, we decided to discard from our calculations those pairs of nodes in which at least one of the nodes changed their opinion by more than 0.05.

*5.2. Tie appearing*

Figure 5 demonstrates the impact of three principal network formation mechanisms – transitivity, preferential attachment, and selectivity (with respect to the age, gender, opinion, and degree attributes) – on the probability of tie appearing. From Figure 5, one can conclude that this quantity goes up with the number of common friends. Further, this covariate has the largest effect on tie creation (see panel A) – having one common link increases at least ten times the chance of becoming friends. Next, in line with previous empirical studies on preferential attachment, we report that vertices with many friends attract more new ties (see panel B). However, the preferential attachment mechanism is in charge only of those ties that appear between vertices that have no common friends (panel C). Instead, if two vertices have one or more common peers, then the greatest chance of tie appearing is achieved if both the nodes have only a few friends (see panels D and E) – a phenomenon that is called anti-preferential attachment (Sendiña-Nadal et al., 2016). In other words, our findings indicate that preferential attachment, triadic closure, and anti-preferential attachment mechanisms are mutually dependent in tie formation processes. We also report that this finding is valid even after controlling for the age, gender, and opinion covariates (see Figures C4–C6 in Appendix C).

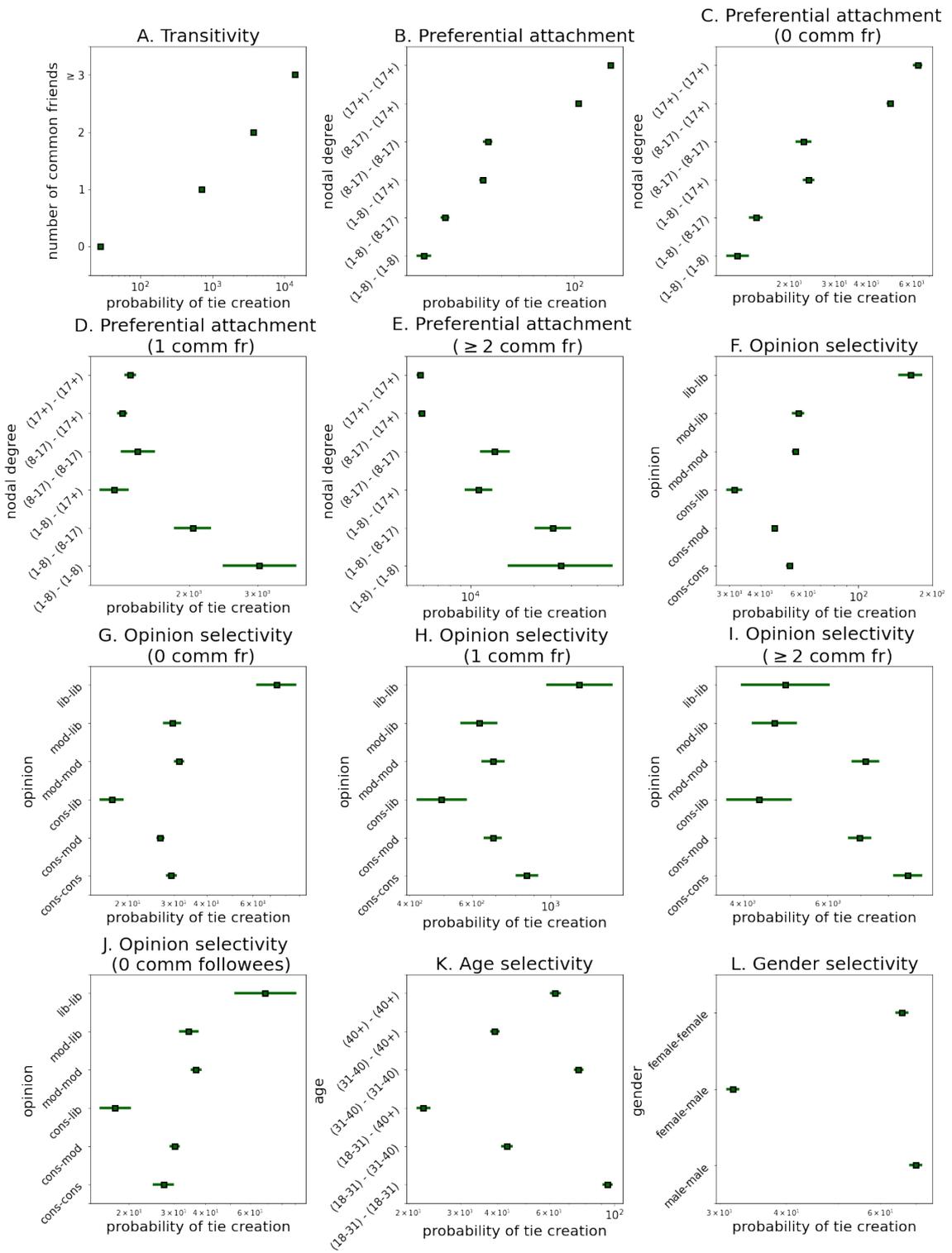

*Figure 5.* The probability of tie appearing as a function of nodal and edge attributes.

Next, Figure 5 is clear evidence of opinion selectivity: for a fixed opinion group, the highest probability of tie creation is achieved if both the nodes belong to this group (see panel F). Instead, the most distant opinion groups (conservatives and liberals) demonstrate the lowest tie creation probability. Interestingly, we revealed quite an interesting difference in tie formation patterns across ideological groups. If two vertices have zero or one common friend, then the pair "lib-lib" features the highest tie creation rate (see panels G or H). In turn, two users that have two or more common friends will have the highest chance of becoming friends if these users espouse conservative opinions (panel I). Importantly, opinion selectivity is observed even if two potential friends have no shared followees (see panel J). And finally, opinion selection remains even after controlling for other covariates (see Figures C7–C9 in Appendix C).

Moreover, Figure 5 demonstrates clear signs of age and gender social selection: individuals from different age or gender groups demonstrate a relatively low tie creation rate (see panels K and L). Further, pairs of younger vertices have more chances to form connections (compare, for example, the pairs "(18-31) – (18-31)" and "(40+) – (40+)").

## 5.3. Tie removing

The next question is what determinants stand behind tie removing. Figure 6 clearly indicates that both the triadic closure and preferential attachment mechanisms contribute to the process of edge destruction in a "reverse fashion": the more common friends two connected individuals have, the less probable it is that they will cut the connection (see panel A). Consequently, high-degree nodes are less frequently involved in deleting ties (even after controlling for the number of common friends – see panels B and C). We also found that the probability of tie removing is higher if users with liberal views are involved (see panel D).

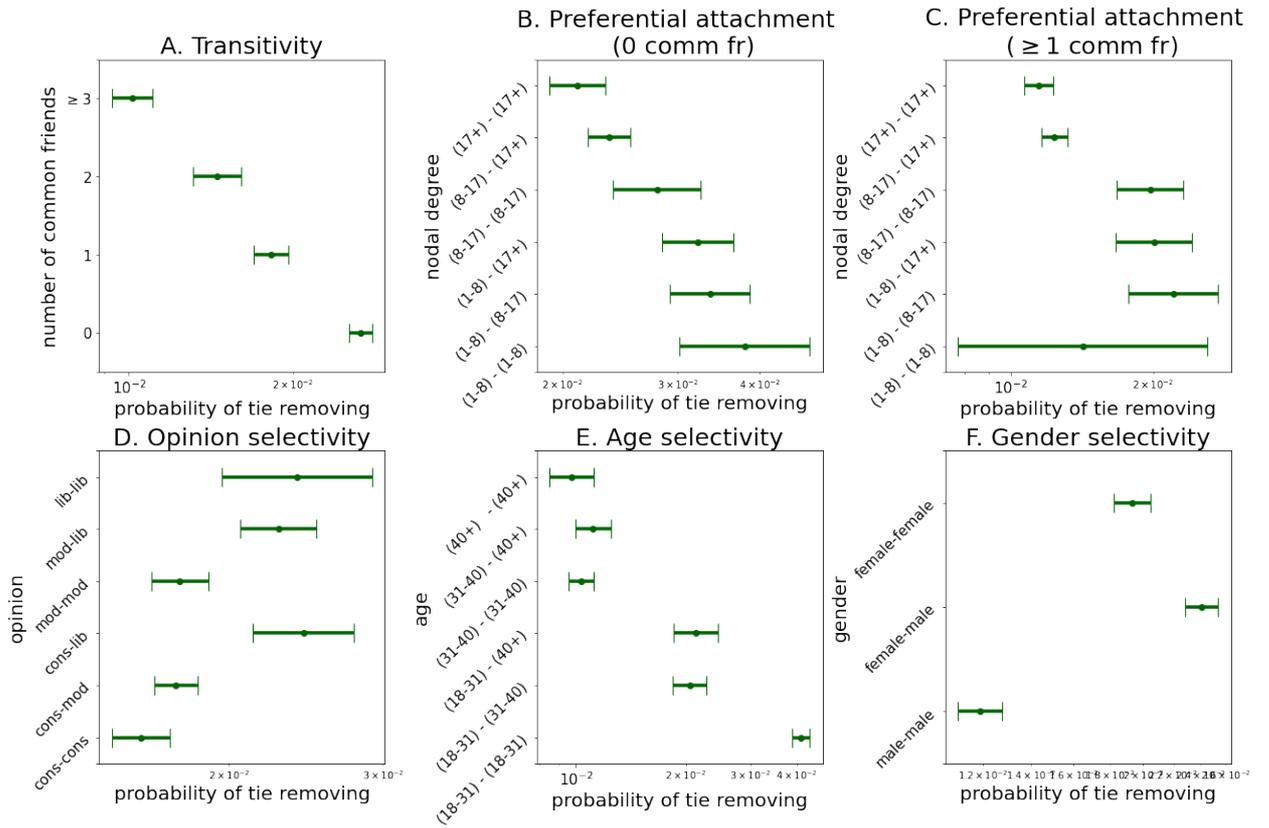

*Figure 6.* The probability of tie removing as a function of node-type and edge-type attributes.

Let us now describe how demographic characteristics affect the tie removing processes. Our results indicate that the gender selectivity factor still has a significant influence on the social system (see Figure 6, panel F), which cannot be said about age selectivity (see panel E). Instead of this, all Figure 6 demonstrates is that younger individuals delete social ties more often. We also report that the connected pairs "male-male" delete their ties less frequently than other pairs (see panel F). Instead, the edge "female-male" displays the highest probability of tie removal. Interestingly, our analysis demonstrates that among the edge configurations "female-female" and "male-male," the second one tends to be more stable.

## 6. Discussion

In this section, we would like to focus on the dynamic patterns of the social system we discovered previously, because its static peculiarities are largely in line with what we expect from social networks. Perhaps only two results of the static analysis are really worth mentioning. First, we revealed a clear asymmetry in the structure of opinion homophily, with radical liberals having a strictly sounder tendency to maintain ideologically homogeneous networks than their conservative counterparts. Further, this structure varies with age and for older individuals, only radical liberals have more connections with those having the same opinions than one could expect by chance, whereas the other opinion groups do not feature such a tendency.

Prior to analyzing and discussing the system's dynamic patterns, it is important to note two important confounding factors that we cannot control but should be carefully accounted for in reflections: (i) the activity of users and (ii) the VK architecture. The first confounder is that different users exhibit varying intensities in online interactions: some users are more active on VK and thus may update their opinions or friend lists more frequently – in particular, this applies to younger individuals who demonstrate higher activity levels online (Subsection 4.2 confirms this). In turn, the factor of the VK architecture can show itself in various ways. First, recommendation systems tend to suggest new friends to users by relying on specific metrics of proximity, for example, the number of common friends, which tend to be a basic driver of recommendations on social networking sites. Next, ranking algorithms change the order of news feed items according to users' preferences and the history of users' actions in the online environment. As a result of such sorting, those pieces of content that may potentially bring any (for example, ideological) discomfort to a user can be hidden from them at the bottom of the news feed. Furthermore, the VK architecture gives users an opportunity to remove the posts published by some of their online friends without deleting connections to these friends.

Now let us go through our findings. First, we discovered that the transitivity and gender selectivity mechanisms have a significant impact on both tie appearing and removing processes: individuals with shared connections and those of the same gender form ties more frequently, and the ties formed between them are more stable. What is more, the triadic closure mechanism seems to have the largest effect on network formation. However, as we have previously mentioned, the effect of transitivity may also stem from VK recommendation systems.

In contrast, the age selectivity factor covers only one side of network formation processes, with no effect on tie removal. We explain this result by noticing that tie creation events include two necessary ingredients: (i) two individuals should meet (no matter, online or offline) and (ii) they should be minded to form a tie (and confirm this tie online). In turn, meeting events are contributed to by (i) shared social contexts and (ii) activities of individuals. The difference in age can be understood as a proxy of shared social contexts: individuals belonging to the same age group have more chances of studying at the same university or having the same job, so they have more chances to meet each other. Though the activity of users affects both the directions of network formation – recall that according to Figures 5 and 6, younger individuals form new connections and delete existing ones more frequently – the influence of the user activity on tie creation is overcompensated by the link between age proximity and the likelihood of meeting, we hypothesize. As a result, we observe a clear effect of age selectivity on the tie creation probability. Instead, Figure 6 indicates that the activity of users dominates the (possible) influence of age selectivity on the tie removing.

Figure 5 also hints that opinion selection is also a driver of tie creation. What is important, this driver is robust against the other covariates, including age. Apparently, the factor of age affects the tie creation processes; we suppose that the observation that liberals form mutual connections more frequently than other pairs is a consequence of the causal relation between age and political opinion – liberals are younger and thus should be more active online. However, even after controlling for age, we end up with the conclusion that the absolute difference in opinions has a negative effect on the tie creation rate (see Figure C7 in Appendix C). It is interesting to note that after controlling for the number of common friends, we revealed that two users with two or more common friends demonstrate a higher probability of tie creation if they are both conservatives. We would like to emphasize that the effect of opinion selection still remains even after considering pairs of vertices with no common followees. To understand why this observation is interesting, one needs to remember that users' opinions were estimated according to their followees. In this regard, one may hypothesize that individuals with close opinions are likely to have shared followees and thus should have more chances to (i) meet each other in the online environment and (ii) become friends due to a shared political identity. However, our results indicate that even individuals with no common followees but those whose followees are of the same political stance feature a higher probability of tie creation.

Our results on the influence of opinion selectivity on tie removing are rather elusive. One can hypothesize that younger individuals are more active on VK and thus ties between them should disappear more often – this is exactly what Figure 6 (panel E) tells us. However, after controlling for the age factor (see Figure C10 in Appendix C), we came to rather mixed evidence. On the one hand, in the zone of relatively small values of $|x_i - x_j|$, not very sound positive relationships between the quantity $|x_i - x_j|$ and the probability of tie removing can be found across all pairwise age combinations. On the other hand, if the difference in opinions becomes too large, then the tie removing probability demonstrates a decrease (see panel A, C, and D on Figure C10). However, this is not always the case (see panel B). These dependencies challenge the idea of selectivity, according to which the probability of tie removing should increase monotonically with the absolute difference in opinions. To some extent, our results can be explained by referring to the peculiarities of the VK architecture (see the beginning of this section).

The most interesting result we have observed is that the transitivity and preferential attachment mechanisms are mutually dependent and do not operate simultaneously: for a given pair of unconnected nodes, if the nodes have no mutual connections, then the probability of tie creation follows the preferential attachment rule and goes up with the total number of nodes' friends. Instead, if the nodes have at least one common friend, then the highest probability of tie appearing is achieved if both the nodes have only a few friends – this mechanism is referred to in the literature as the anti-preferential attachment mechanism. Interestingly, this peculiarity does not work in reverse: the probability of tie removing is lower if both connected nodes have many social connections, regardless of the number of common peers.

First, this finding specifies the circumstances under which the preferential and anti-preferential attachment mechanisms appear. This knowledge can be incorporated into network formation models to achieve more accuracy. For example, in the Ref. (Sendiña-Nadal et al., 2016), the authors attempted to introduce these mechanisms into a network formation model by dividing the nodes into two populations (followers and leaders), whereby the first one (followers) follows the preferential attachment mechanism, whereas the second one (leaders) follows anti-preferential attachment – this allow leaders to get more popular in the network. However, our result hints at a different way to apply preferential and anti-preferential attachment, with a special control for triadic closure.

In a nutshell, this result can be reformulated as follows: two high-degree nodes that have no common friends or just a few of them have a relatively high chance of becoming friends (compared to low-degree nodes with the same number of common friends), whereas two high-degree nodes that have more common friends establish ties less frequently, than low-degree vertices with the same number of common peers. In this formulation, we use the words like "just a few of them" just because we do not know exactly how many common friends a given pair of users have on VK – we are working with the sample only, not with the entire network. The number of common friends is a metric that can be understood as an indicator that two nodes pertain to the same community: those who have no common friends or only a few of them are likely members of different social communities, whereas those who have many common peers likely belong to the same community. In turn, high-degree nodes can be interpreted as community leaders. Of course, social communities are complex structures whereby two or more leaders could coexist and maintain different types of relationships with each other, such as competition, or alliance. If we know that two leaders within the same community are not connected at this moment, then one can hypothesize that they are competitors in some way. In this regard, there is a relatively low likelihood that they will become friends in the near future, so the structural hole will remain. Instead, if we consider two leaders that pertain to different communities, then they are unlikely to have reasons for a conflict, compared to the situation when they belong to the same community. Further, these two leaders have more chances to become friends than if we would consider other representatives of the communities just because the leaders are more popular. In contrast, low-degree nodes within the same community are unlikely to have points of hostility, compared to if they were community leaders. In turn, if two leaders from the same community are connected, then one can conclude that they are not competitors so they both will have losses, as well as the community at all, if they break the connection. Because of this, the anti-preferential attachment mechanism does not affect tie removal in the case of individuals that have common friends. Figure 7 schematically presents our ideas.

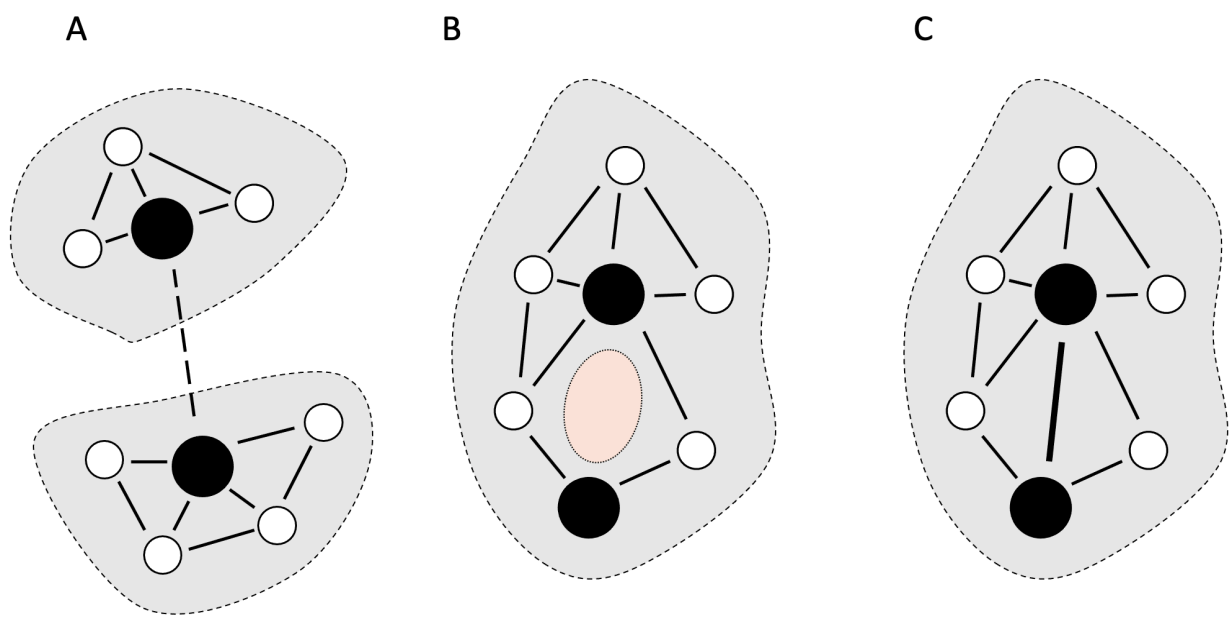

*Figure 7*. (Panel A) On this panel, two communities are presented (emphasized by gray areas), each of them has a leader (big black circle). Because the community leaders are unlikely to have reasons for a conflict (their zones of interests do not intersect), there is a relatively high chance that they will become friends (dashed edge). (Panel B) Here, we consider only one community that has two leaders, each of them attempts to become a sole champion. Due to this competition, the leaders are not connected, so one can observe a stable structural hole (orange ellipsis). (Panel

C) In this configuration, two community leaders are some sort of ally. Their friendship (bold edge) is a relatively stable formation that benefits not only the leaders, but the entire community.

We believe that our results can be useful in further studies on social networks. In particular, it would be interesting to test our finding on the interplay between triadic closure, preferential attachment, and anti-preferential attachment in network formation models. Furthermore, our results on opinion selectivity and the structure of different types of homophily would be useful in studies of social influence processes. Not to mention the dataset, which can be used in studying the coevolution of opinions and social ties.

## 7. Limitations and Generalizability

Our research is not without limitations. The first one relates to the object of study – online networks store only one facet of communication processes, so an important part of information was likely hidden from our analysis. Further, data from online networks are typically incomplete – for example, we had no opportunity to control for the factor of the shared residence, which plays an important role in network formation (Lewis et al., 2012). The only geographical restriction we applied was to live in the same city, which was chosen as a large city (the capital of a region), sufficiently distant from other huge settlements – this approach facilitates isolation of the social system and "completeness" of the extracted network (Flache et al., 2017).

Despite this, we believe that our sampling strategy was as fair as possible. However, it had one more weakness. To facilitate the precision of the opinion estimation procedure, we excluded from the sample all users that had too little or too many followees. Because users' activity patterns typically correlate, one should expect that individuals that have only a few followees will also have only a small number of online friends (our correlation analysis confirmed this). Because of this, one can suppose that by excluding individuals that have little to no followees, we disposed of the periphery of the social network. On the one hand, such a strategy facilitated purity of the sample

– because the online bots typically occupy the periphery of online networks (González-Bailón & De Domenico, 2021)) and have unusual numbers of followees. On the other hand, there was a high probability of missing newly registered users that only started their online presence and did not have large ego-networks. Such "fresh" accounts could feature completely different behavior patterns and thus affect analysis.

An important limitation of our study stems from the architecture of the online network VK, which tends to facilitate communication between like-minded individuals or recommend new friends according to specific user-similarity measures, such as the number of common friends – see the beginning of the previous section for details.

We should also say that our analysis focused on the dynamics of social ties between rather ordinary (native) users, whereas popular users (bloggers) were beyond the scope of our vision. This step was motivated by our opinion estimation procedure, which employs information of users' followees (including bloggers) as independent variables. Here, it is worth saying that users' opinions were likely estimated with errors, which could also affect our conclusions regarding the effect of opinion selection.

To test the generalizability of our findings, we replicated our analysis two more times using two independently developed samples of users, built according to the same rules. This additional investigation (see Online Supplementary Materials for details) revealed the robustness of our results. Besides, as an additional robustness check, we used different threshold values in making age, opinion, and nodal degree stratifications. All results were virtually the same. Of course, this does not necessarily mean that our findings can be generalized to other empirical settings. For example, different online platforms provide different organizations of communication environments, so our results may not be applicable for some of them.

## 8. Data availability

JupiterHub was used to perform data processing, analysis, and visualization using the Python 3 programming language. All of the data, codes, and other supporting information can be found at https://doi.org/10.7910/DVN/D7RUJN (Online Supplementary Materials). Before starting to investigate the code, please read the file "Manual.pdf."

## 9. Authors' contributions

I.K. conceived and designed the research, processed the data, analyzed the results, and wrote and revised the manuscript.

E.R. and V.L. obtained raw data from the online network VKontakte, reviewed, and edited the manuscript.

## 10. Funding

The research was supported by a grant of the Russian Science Foundation (project no. 22- 71- 00075).

# Appendix A. Assortativity coefficient definition

If we say that a network $G$ is homophilic with respect to some attribute $x$, it means that we take a look at $G$ at a fixed time moment $t$ (that is, observe its snapshot $G(t)$) and see that $G(t)$'s neighboring nodes tend to be similar in terms of this attribute $x$. A possible approach to understanding whether $G(t)$ is homophilic with respect to the attribute $x$ is to calculate the assortativity coefficient:

$$C(A,x) = \frac{\sum_{i,j}\left(a_{ij} - \frac{k_i k_j}{2q}\right)x_i x_j}{\sum_{i,j}\left(k_i \delta_{ij} - \frac{k_i k_j}{2q}\right)x_i x_j} \in [-1,1],$$

where the adjacency matrix $A = [a_{ij}]$ formalizes the edges in the network $G(t)$, $k_i$ stands for the degree of node $i$ ($k_i = \sum_j a_{ij}$), $x_i$ represents the node $i$'s attribute, $\delta_{ij}$ is the Kronecker function, and $q = \sum_{i,j} a_{ij}/2$ signifies the number of edges in the network. There are some technical differences in calculating the assortativity coefficient depending on whether the focal attribute $x$ is numeric or not. Note that the formula above is applicable in situations when $x$ is numeric. For more details, see Ref. (M. Newman, 2018) and official documentation of the Networkx package on https://networkx.org/documentation/stable/reference/algorithms/assortativity.html.

# Appendix B. Opinion estimation approach

We estimated users' opinions on a political scale by leveraging the information of their followees. In particular, we apply the approach elaborated in Ref. (Kozitsin et al., 2020). Within this approach, a few seed information sources (bloggers and public pages) were chosen as representatives of the main politicians of the Russian political landscape. Followers of these accounts (there were 63,538 of such users) are assumed to be supporters of the corresponding political figures and can be understood as labeled instances, following the language adopted in the machine learning theory. Of course, some of these users were identified as followers of several seed political accounts at once, but the population of such users was extremely tiny (~100 users).

Next, we identified the set $J$ of all public pages and bloggers followed by these users – it was composed by $N_s = 3{,}692{,}970$ accounts. It is worth noting that this set included the seed accounts. All accounts from the set $J$ were used to develop a feature space, in which each user $i$ was characterized by a string

$$D_i = \begin{bmatrix} d_i^1 & \ldots & d_i^{N_s} \end{bmatrix}$$

that represents which information sources this user was subscribed to (dummy-encoding). In this string, the element $d_i^k \in \{0,1\}$ signifies whether the user followed the $k$-th information source.

Following that, a logit classifier (chosen as the best classifier capable of producing interpretable output in the form of posterior probabilities) was applied to the labeled data (which had previously been projected into a 100-dimension space by the truncated singular value decomposition transformer). During training, this model pursues finding those logit regression coefficients that ensure the lowest value of a loss function. Note that without dimensionality reduction, the resulting coefficients would demonstrate how important this or that information source is in the classification problem.

Initially, we solved a multiclass classification problem whereby the classes corresponded to the fixed political figures. However, pilots revealed that the most precise classification can be achieved if considering only two classes – (i) supporters of President Putin or the leaders of the Russian systemic opposition (in the main part of the paper we call them conservatives) and (ii) backers of the Russian non-systemic opposition (we call them liberals). Because of this issue, we decided to focus on the binary classification problem. To avoid overfitting (caused by seed accounts, which draw attention of the classifier to themselves because they are obvious markers of true classes), we used artificially created accounts that were obtained as copies of labeled accounts with erased subscriptions to seed information sources. These "truncated" observations were used in training, and, importantly, only such accounts were used in testing the classifier. It allowed us to get a more strict, fair, and unbiased estimation of classification quality (perhaps, too strict). Because the labeled data was slightly unbalanced, we used the balanced accuracy as a baseline metric. We managed to achieve the quality of classification of 0.76 on a test sample. The resulting confusion matrix is depicted in Figure B1.

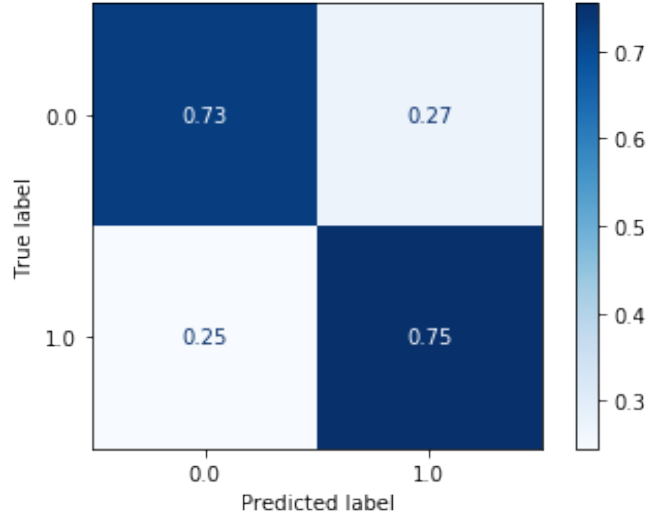

*Figure B1*. This confusion matrix shows the quality of classification on the test sample (the quality metric – balanced accuracy). Class 0 - supporters of President Putin or the leaders of the systemic opposition, class 1 - backers of the non-systemic opposition.

For a given user $i$ that at time moment $t$ is described by a string $D_i^T(t) = [d_i^{T,1} \quad ... \quad d_i^{T,100}]$, the model produces the following estimation:

$$x_i(t) = \frac{1}{1 + e^{-\beta_0 + \sum_{k=1}^{100} \beta_k d_i^{T,k}}} \in [0,1],$$

where $\beta_0, \beta_1, ..., \beta_{100}$ are regression coefficients estimated during the training procedure. Note that in the event we do not perform the dimensionality production procedure, the opinion estimation is given by

$$x_i(t) = \frac{1}{1 + e^{-\beta_0 + \sum_{k=1}^{N_s} \beta_k d_i^k}} \in [0,1], \qquad (B1)$$

where the quantities $d_i^k$ are the components of the dummy vector $D_i(t)$.

The quantity $x_i(t)$ can be understood as the conditional probability of being a supporter of the non-systemic opposition. In turn, coefficients $\beta_1, ..., \beta_{N_s}$ in (B1) demonstrate how the corresponding information sources contribute to classification. Without dimensionality reduction, if $\beta_k$ is large and positive, it means that the $k$-th information source is valuable in classification and a subscription to this account is a clear signal of being a member of the non-systemic opposition. As an additional check, we trained the classifier without the dimensionality reduction step, chose 200 largest in absolute value coefficients of the trained classifier, and then verified the corresponding information sources. The majority of them were dedicated to political issues to some extent, and, importantly, their political bias matched the signs of the coefficients.

Then, the model trained on the whole labeled dataset was applied to the sample users. Because we had two snapshots of users' followees, we obtained two opinion estimations for each user. In this regard, the changes in users' opinions in our data stemmed from the changes in their followee lists.

# Appendix C. Auxiliary Figures and Tables

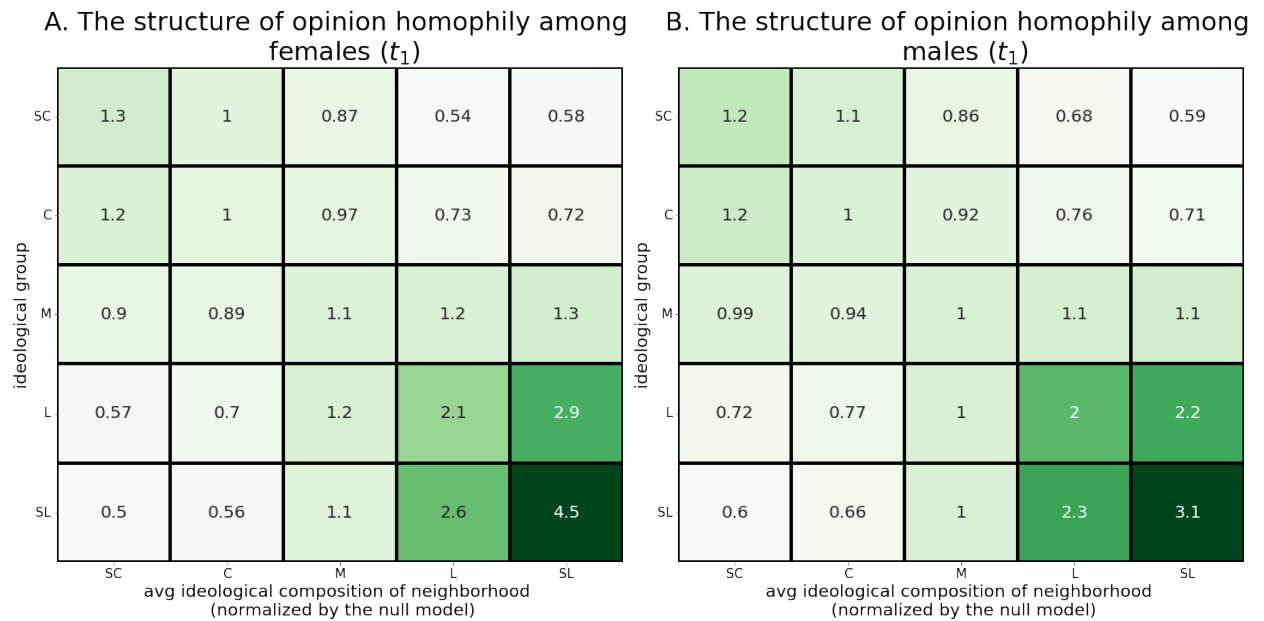

*Figure C1*. The structure of opinion homophily at time $t_1$ for females (left panel) and males (right panel).

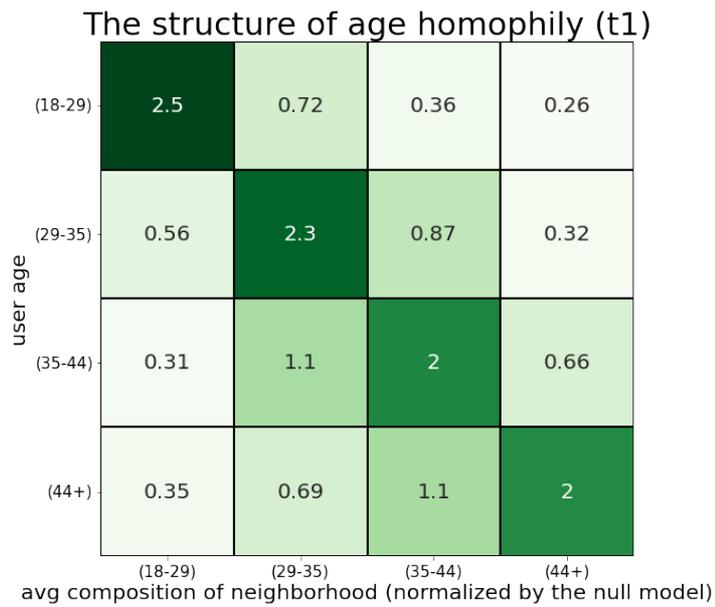

*Figure C2.* The structure of age homophily at time $t_1$.

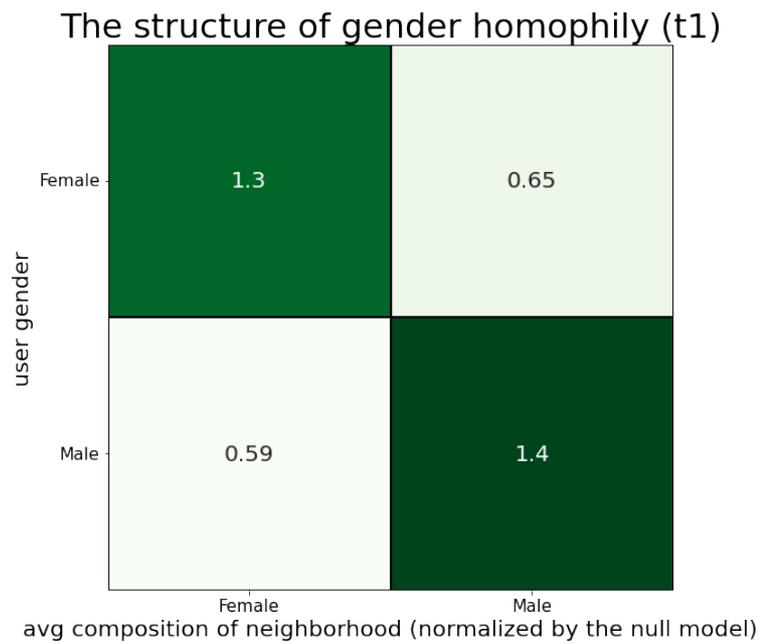

*Figure C3.* The structure of gender homophily at time $t_1$.

Table C1

Users' characteristics across ideological groups. In each cell, the first value represents the median, the second one – the mean.

|  | Strong Conservatives | Conservatives | Moderates | Liberals | Strong Liberals |
| --- | --- | --- | --- | --- | --- |
| Number of friends | 13<br>16.55 | 11<br>14.92 | 12<br>16.94 | 13<br>18.66 | 15<br>20.69 |
| Number of new friends | 1<br>1.41 | 1<br>1.29 | 1<br>1.63 | 1<br>1.6 | 1<br>1.53 |
| Number of deleted friends | 0<br>0.31 | 0<br>0.26 | 0<br>0.34 | 0<br>0.49 | 0<br>0.57 |
| Number of followees | 93<br>96.28 | 45<br>54.6 | 29<br>40.34 | 51<br>61.49 | 73<br>80.25 |
| Number of new followees | 6<br>8.97 | 4<br>7.05 | 3<br>6.15 | 5<br>7.73 | 7<br>9.69 |
| Number of deleted followees | 4<br>6.26 | 2<br>4.25 | 2<br>3.91 | 3<br>6.62 | 5<br>9.44 |
| Age | 36<br>38.13 | 37<br>39.79 | 35<br>36.77 | 28<br>30.11 | 26<br>28.67 |
| Gender (1 – female, 2 – male) | 2<br>1.54 | 1<br>1.44 | 1<br>1.49 | 2<br>1.51 | 2<br>1.56 |

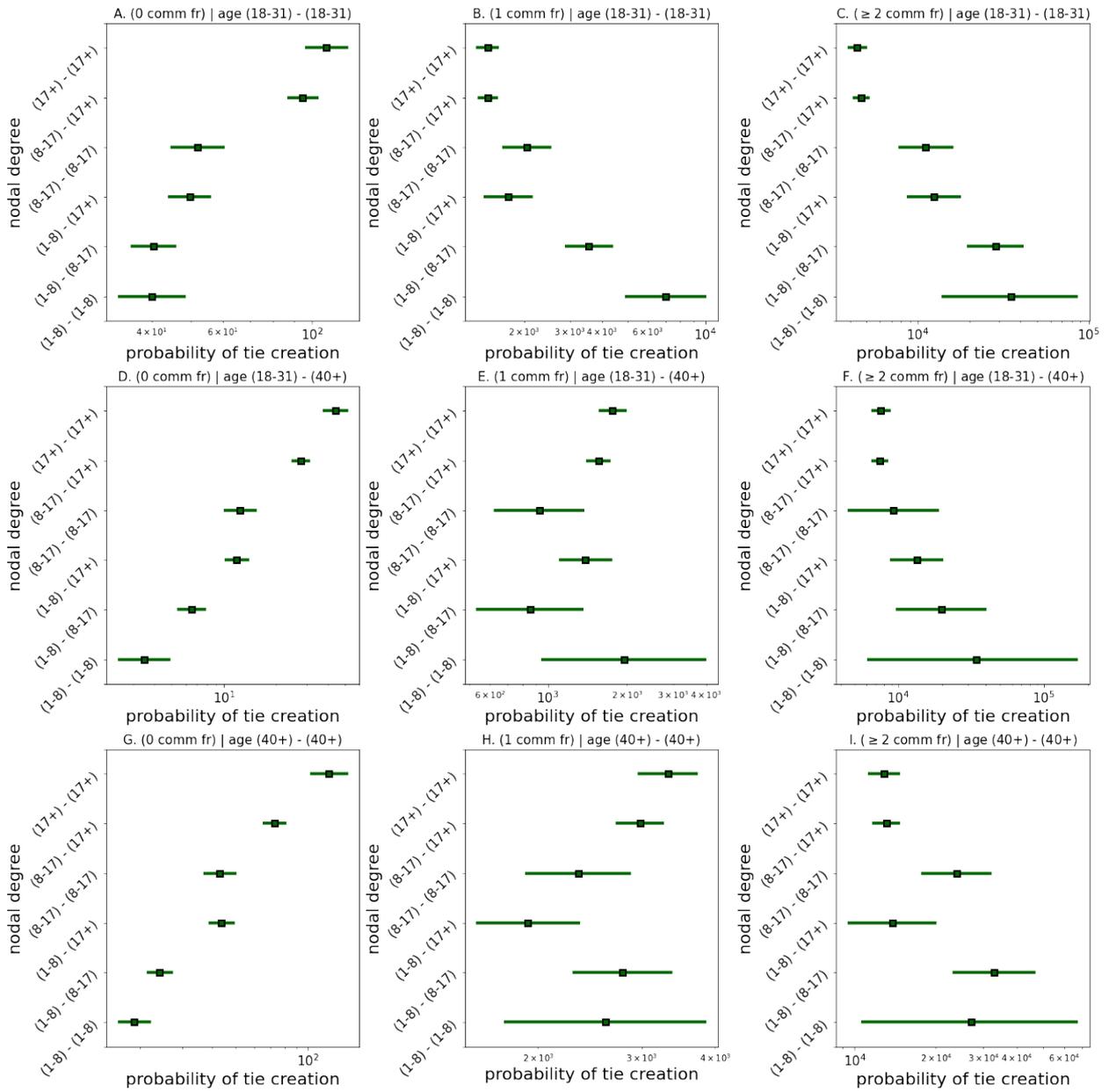

*Figure C4.* The probability of tie creation as a function of nodes' degrees across different pairwise age combinations subject to the number of common friends.

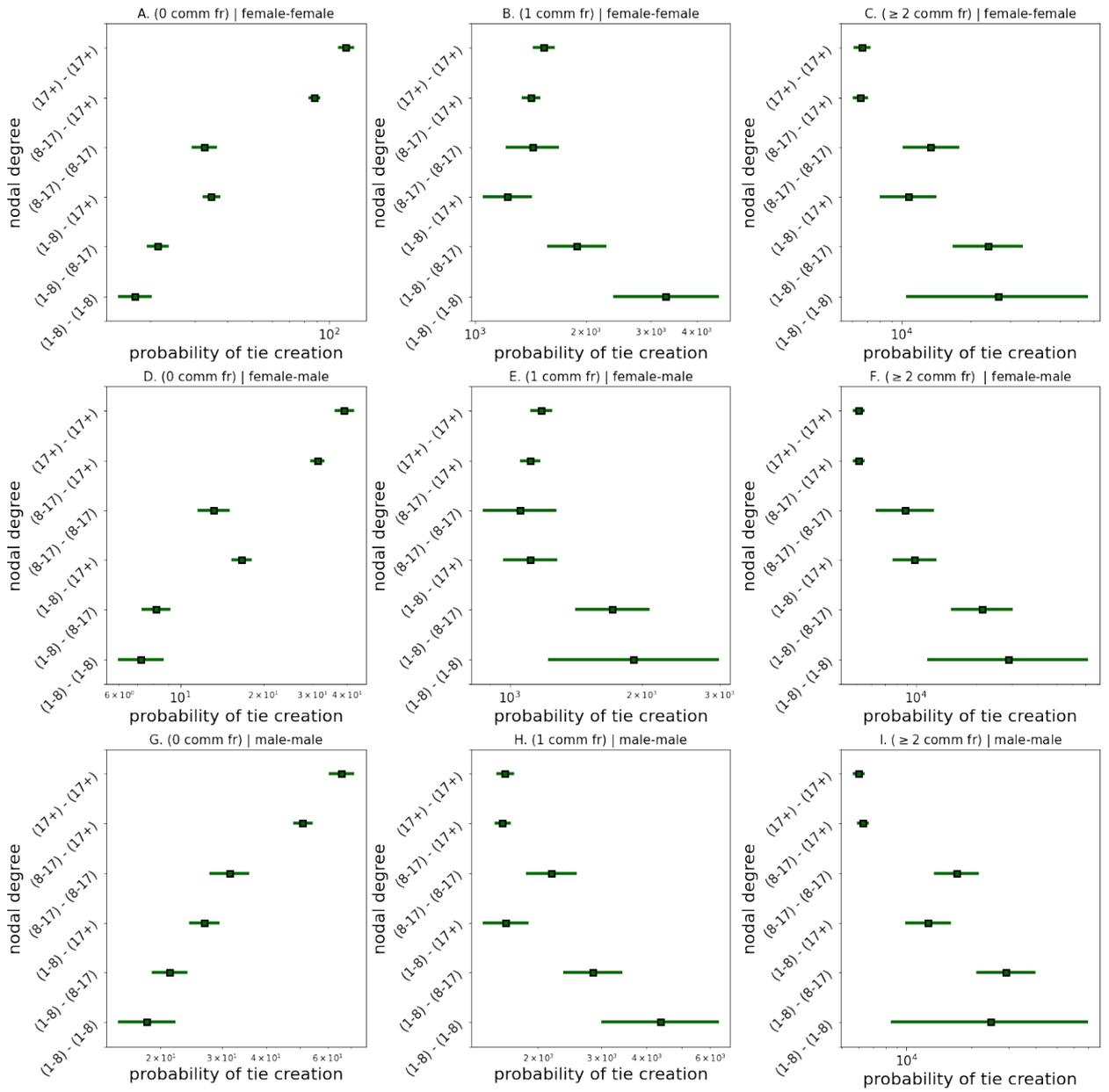

*Figure C5.* The probability of tie creation as a function of nodes' degrees across different pairwise gender combinations subject to the number of common friends.

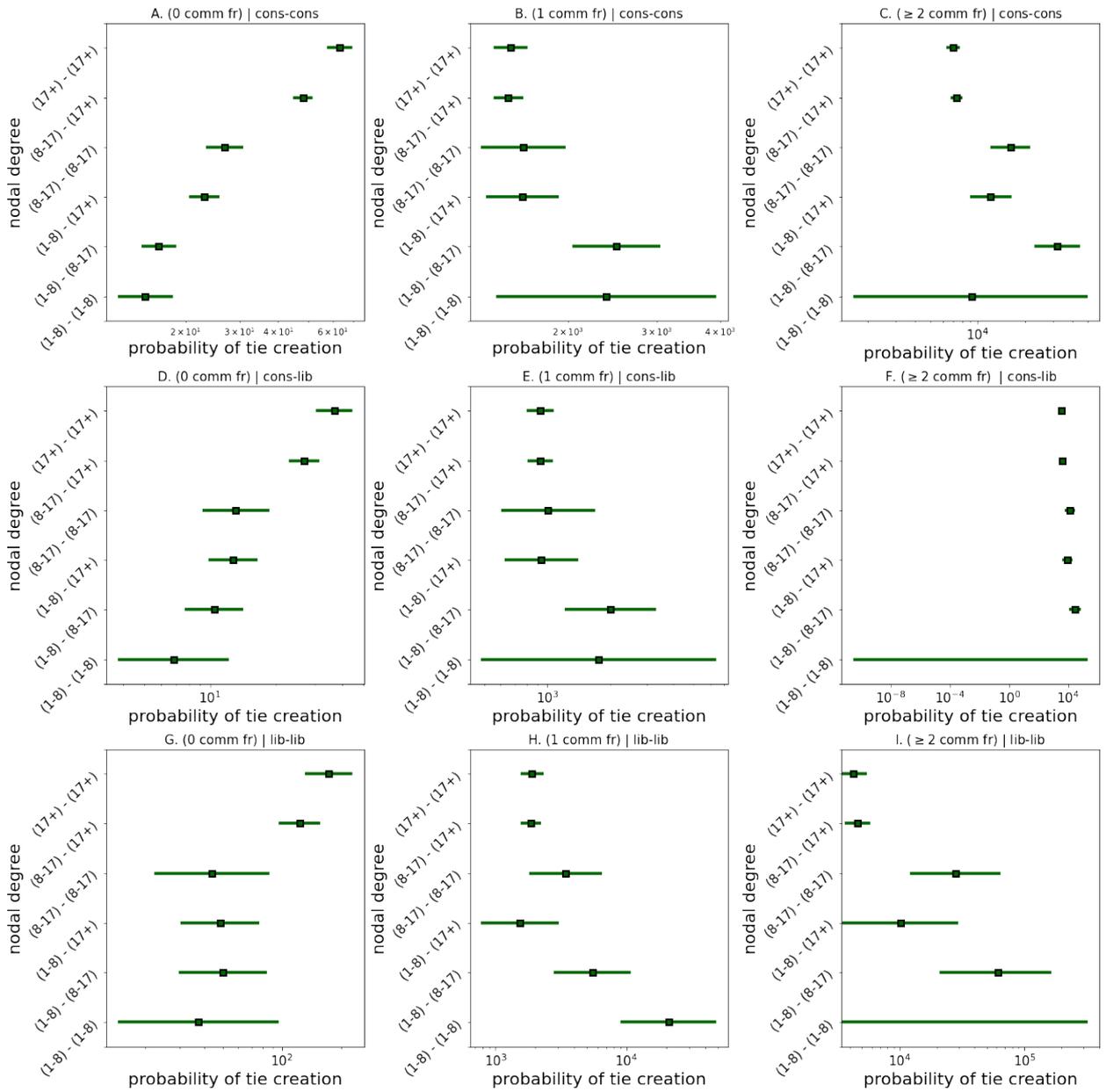

*Figure C6.* The probability of tie creation as a function of nodes' degrees across different pairwise opinion combinations subject to the number of common friends.

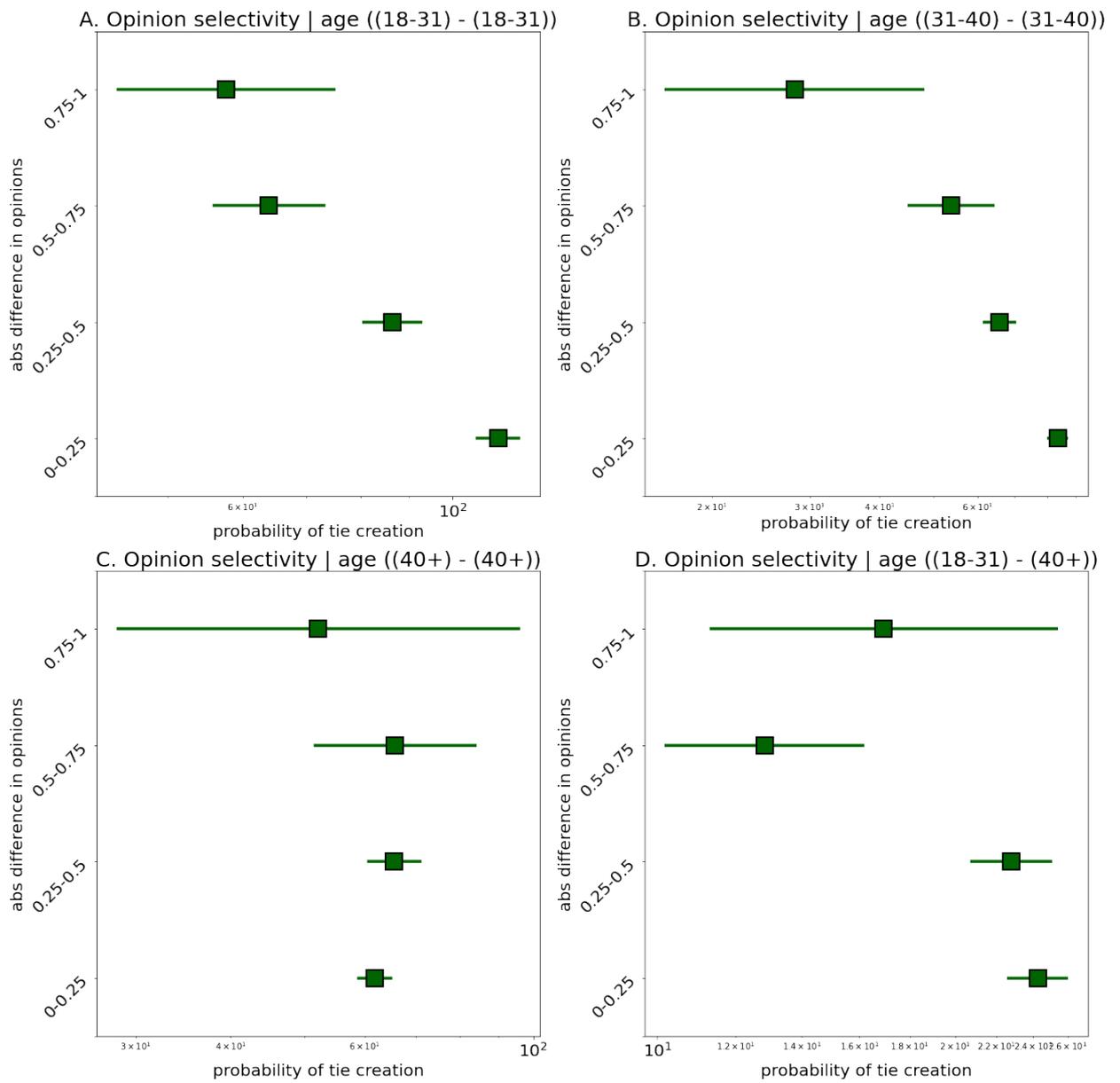

*Figure C7.* The probability of tie appearing as a function of the absolute difference in opinions across different pairwise age combinations.

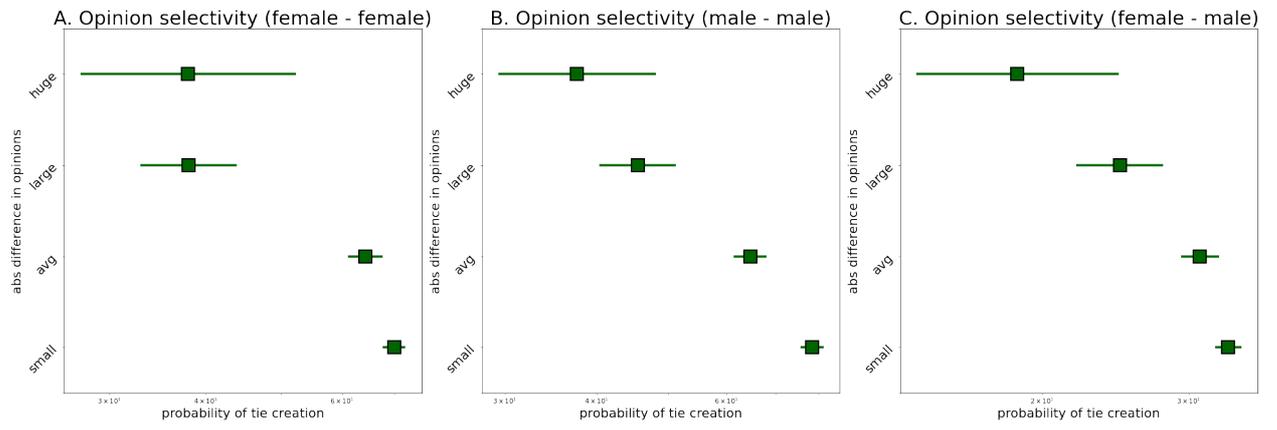

*Figure C8*. The probability of tie appearing as a function of the absolute difference in opinions across different pairwise gender combinations.

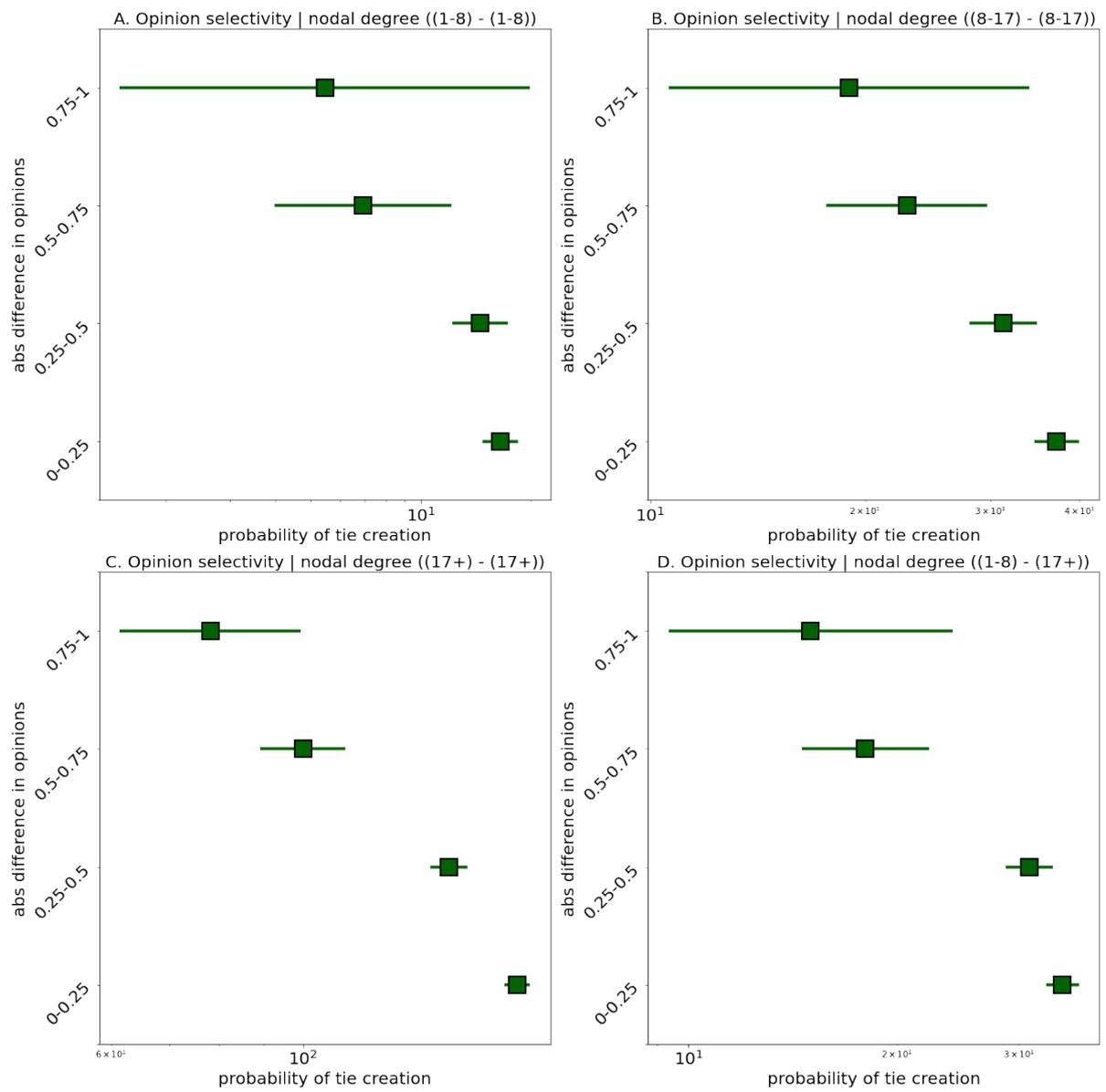

*Figure C9.* The probability of tie appearing as a function of the absolute difference in opinions across different pairwise node degree combinations.

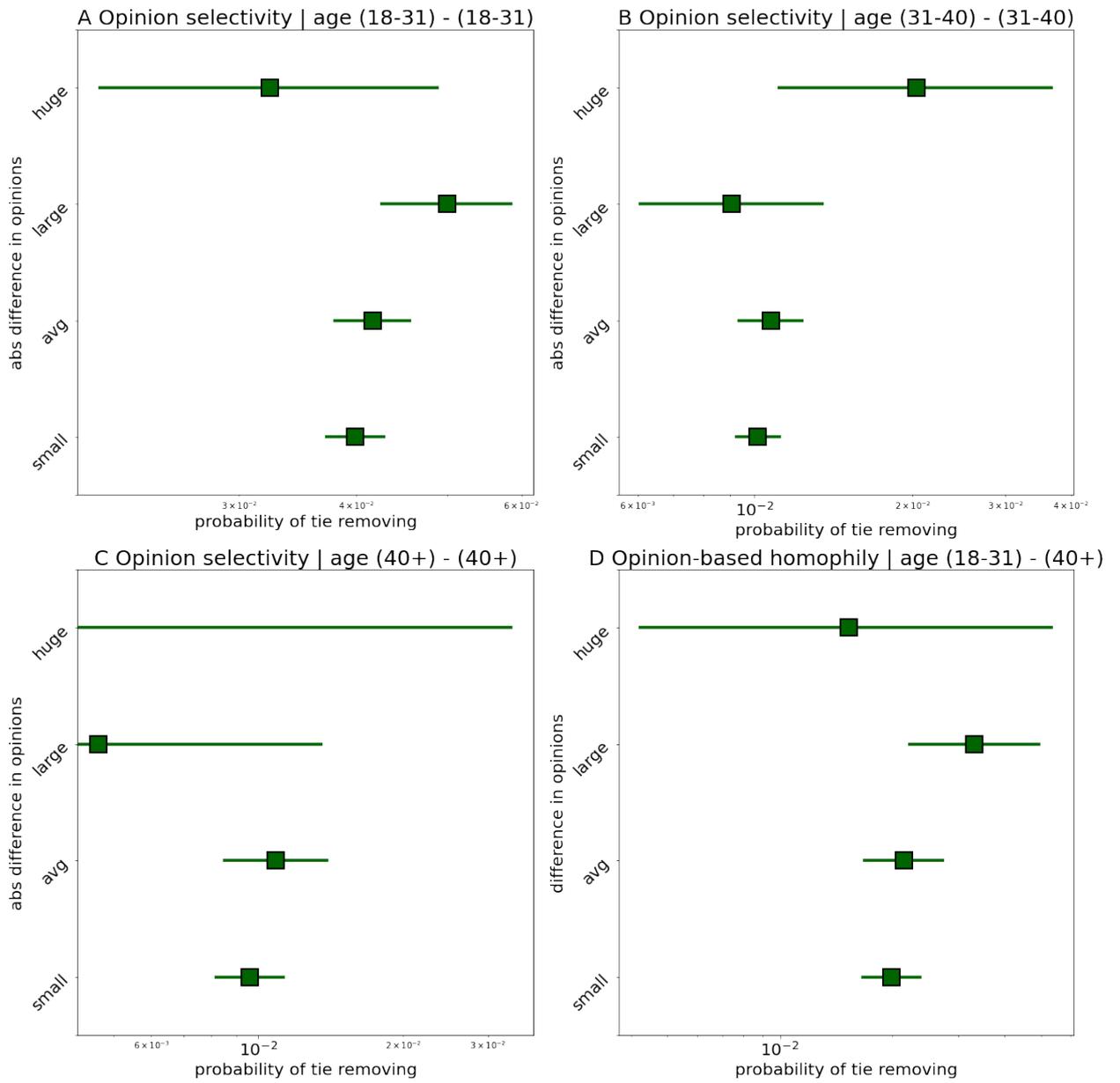

*Figure C10*. The probability of tie removing as a function of the absolute difference in opinions across different pairwise age combinations.